\documentclass[12pt, letter, english, twoside]{myarticle}

\usepackage{graphicx}
\usepackage{fancyhdr}   
\usepackage{cite} 
\usepackage{amsmath, subeqnarray}
\usepackage{indentfirst}

\DeclareMathOperator{\trace}{Tr}
\newcommand{\eAA}{\epsilon_{AA}^{(1)}}
\newcommand{\eAB}{\epsilon_{AB}^{(1)}}
\newcommand{\eBB}{\epsilon_{BB}^{(1)}}
\newcommand{\eAAA}{\widetilde{\epsilon}_{AAA}^{(1)}}
\newcommand{\eAAB}{\widetilde{\epsilon}_{AAB}^{(1)}}
\newcommand{\eABB}{\widetilde{\epsilon}_{ABB}^{(1)}}
\newcommand{\eBBB}{\widetilde{\epsilon}_{BBB}^{(1)}}
\newcommand{\eAAAA}{\widetilde{\epsilon}_{AAAA}^{(1)}}
\newcommand{\eAAAB}{\widetilde{\epsilon}_{AAAB}^{(1)}}
\newcommand{\eAABB}{\widetilde{\epsilon}_{AABB}^{(1)}}
\newcommand{\eABBB}{\widetilde{\epsilon}_{ABBB}^{(1)}}
\newcommand{\eBBBB}{\widetilde{\epsilon}_{BBBB}^{(1)}}

\newcommand{\cf}{\textit{cf.}{ }}
\newcommand{\etal}{\textit{et al.}{ }}
\newcommand{\ie}{\textit{i.e.}{ }}

\renewcommand{\headrulewidth}{0.pt}
\newcommand{\myheading}[1]{ \footnotesize{\textbf{#1}} }
\newcommand{\figcaption}[1]{\caption{\footnotesize{#1}}}
\newenvironment{acknowledgments}{\section*{ACKNOWLEDGMENTS}}{}

\usepackage{dcolumn}
\newcolumntype{.}[1]{D{.}{.}{#1}}

\begin{document}
\renewcommand{\headrulewidth}{0.pt}

\thispagestyle{plain}

\hfill \\
\hfill \\
\hfill \\
\hfill \\
\hfill \\
\hfill \\
\hfill \\
\hfill \\
\hfill \\
\hfill \\
\textbf{MONTE CARLO STUDY \\
OF THE PRECIPITATION KINETICS \\
OF Al$_3$Zr IN Al-Zr} \\
\hfill \\
\hfill \\
\hfill \\
\begin{minipage}{0.99in}
\hfill
\end{minipage}
\begin{minipage}{5.125in}
Emmanuel Clouet$^\textrm{1,2}$ and Maylise Nastar$^\textrm{2}$ \\
\hfill \\
\begin{minipage}[t]{0.125in}
$^\textrm{1}$ 
\end{minipage}
\begin{minipage}[t]{5in}
Pechiney, Centre de Recherches de Voreppe \\
BP~27\\
38341 Voreppe cedex\\
France\\
\end{minipage}

\begin{minipage}[t]{0.125in}
$^\textrm{2}$ 
\end{minipage}
\begin{minipage}[t]{5in}
Service de Recherches de M\'etallurgie Physique\\
 CEA/Saclay B\^at. 520 \\
 91191 Gif-sur-Yvette \\
 France
\end{minipage}
\end{minipage}
\hfill \\
\hfill \\
\section{INTRODUCTION}

Precipitation kinetics in alloys, like spinodal decomposition,
nucleation and growth, or phase ordering, are now often studied at
an atomistic scale using Monte Carlo simulations. So as to be able
to reproduce the different kinetic behaviors during these
transformations, one needs to adopt a realistic description of the
diffusion. Therefore it is better to use a vacancy-diffusion
mechanism than a direct atom exchange mechanism. It is then
possible to explain why different kinetic pathways are observed.
For instance, the vacancy diffusion mechanism can predict the
importance of monomer diffusion with respect to the diffusion of
small clusters \cite{SOI00,LEB02,ATH00,ROU01}. This leads to a
difference in the cluster size distribution during precipitation
\cite{LEB02} and determines the coarsening mechanism
(evaporation-condensation or coagulation) \cite{ATH00,ROU01}.
One can predict too the slowdown of precipitation kinetics by vacancy
trapping due to the addition of a third component impurity
\cite{SOI00}. Finally, different interactions of solute atoms with
vacancy lead to a difference of precipitate~/ matrix interface morphology
during the kinetic pathway, the interface being diffuse for a
repulsion between vacancy and solute atoms and sharp for an
attraction \cite{ROU01}.

One drawback of kinetic Monte Carlo simulations using
vacancy-diffusion mechanism is that they limit themselves to pair
interactions to describe configurational energy of alloys.
Multisite interactions including more than two lattice sites are
necessary if one wants to fully reproduce the thermodynamics of a
system \cite{DUCASTELLE,  LU91}. These interactions reflect
dependence of bonds with their local environment and as a
consequence break the symmetry imposed by pair interactions on
phase diagram. It is interesting to notice that in Calphad
approach \cite{CALPHAD} one naturally considers such interactions
by describing formation energy of solid solutions with
Redlich-Kister polynomials, and that coefficients of these
polynomials can be mapped onto an Ising model to give effective
interactions including more than two lattice sites
\cite{DES97,DES98}. Moreover these interactions allow one to
understand shapes of precipitates in alloys \cite{MUL01} and can
lead to a prediction of coherent interface energy in really good
agreement with ab-initio calculations performed on supercells
\cite{AST98}. Nevertheless, to study kinetics in Monte Carlo
simulations with such interactions, one usually uses a direct atom
exchange mechanism \cite{MUL02}, and thus looses all kinetic
effects due to vacancy-diffusion mechanism.

We incorporate these multisite interactions in a kinetic model
using a vacancy-diffusion mechanism to study precipitation kinetics
of Al$_3$Zr in Al-Zr solid solution.
For small supersaturation in zirconium of the aluminum solid
solution, it has been experimentally observed that Al$_3$Zr
precipitates are in the metastable L1$_2$ structure
\cite{RYU69,ROB01,NES72}.  These precipitates are found to have
mainly spherical shape (diameter $\sim 10-20$~nm), as well as
rod-like shape \cite{RYU69}. For supersaturation higher than the
peritectic concentration, nucleation is homogeneous and
precipitates are coherent with the matrix \cite{RYU69,ROB01}. With
prolonged heat treatment, if the temperature is high enough, the
metastable L1$_2$ structure can transform to the stable one
DO$_{23}$. Using a phase field method, Proville and Finel
\cite{PRO01} modelled these two steps of the precipitation, \ie the
transient nucleation of the L1$_2$ structure and the
transformation to the DO$_{23}$ structure.
In this work, we only focus on the precipitation first stage,
where Al$_3$Zr precipitates have the L1$_2$ structure
and are coherent with the matrix.

We first use ab-initio calculations to fit a generalized Ising model
describing thermodynamics of Al-Zr system.
We then extend description of the configurational energy of
the binary Al-Zr to the one of the ternary Al-Zr-Vacancy system
and adopt a vacancy-atom exchange mechanism to describe kinetics.
This atomistic model is used in Monte Carlo simulations
to study diffusion in the Al-Zr solid solution
as well as precipitation kinetics of Al$_3$Zr.
We mainly focus our study on detecting any influence of multisite interactions on kinetics.

\section{THERMODYNAMICS OF Al-Zr BINARY}

\subsection{Ab inito calculations}

We use the  full-potential linear-muffin-tin-orbital (FP-LMTO) method
\cite{AND75,MET88,MET89} to calculate formation energies of
different compounds in the Al-Zr binary system, all based on a fcc
lattice.
Details of ab initio calculation can be found in appendix \ref{appendix_abinitio}.
They are the same as in our previous work \cite{CLO02},
except the fact that we use the generalized gradient approximation (GGA) instead of the
local density approximation (LDA) for the exchange-correlation functional.

The use of GGA for the exchange correlation energy leads to a
slightly better description of the Al-Zr system. The approximation
does not fail to predict phase stability of pure Zr \cite{JOM98}
as LDA does: if one does not include generalized-gradient
corrections, the stable structure at 0~K for Zr is found to be the
$\omega$ one (hexagonal with 3 atoms per unit cell) and not the
hcp structure.

Another change depending on the approximation used for the
exchange-correlation functional is that formation enthalpies
obtained with GGA for the different Al-Zr compounds are a little
bit lower (a few percent) than with LDA. For the DO$_{23}$
structure of Al$_3$Zr (table \ref{Al3Zr_calcul}), GGA predicts a
formation energy which perfectly reproduces the one measured by
calorimetry \cite{MES93}: including generalized-gradient
corrections has improved the agreement. The energy of
transformation from the L1$_2$ to the DO$_{23}$ structure, $\Delta
E=-23$ meV/atom, agrees really well too with the experimental one
measured by Desh \etal \cite{DES91}, but this was already true
with LDA. Gradient corrections have improved the agreement for the
equilibrium volumes too: with the LDA, they were too low compared
to the available experimental ones. Considering the values of the
relaxed equilibrium parameters, shape of the unit cell and atomic
positions, no change is observed according to the approximation
used, both LDA and GGA being in good agreement with measured
parameters.

\begin{table}[!hbt]
\caption{Calculated equilibrium volumes $V_0$, $c'/a$ ratios
($c'=c/2$ for the DO$_{22}$ phase and $c'=c/4$ for the DO$_{23}$
phase), atomic displacements (normalized by $a$), and energies of
formation for Al$_3$Zr compared to experimental data.}
\label{Al3Zr_calcul}
\begin{minipage}{\textwidth}
\renewcommand{\footnoterule}{}
\begin{center}\begin{tabular}{llccccc}
\hline
&& $V_0$         & $c'/a$     & Atomic       & $\Delta E$\\
&& (\AA$^3$/atom)&          &  displacements & (eV/atom)\\
\hline
L1$_2$    & GGA\footnote{FP-LMTO calculations (present work)} & 16.89 &    &   & $-0.478$ \\
&                LDA\footnote{FP-LMTO calculations\cite{CLO02}} & 16.12 &    &   & $-0.524$ \\
\\
DO$_{22}$ & GGA$^a$& 17.40 & 1.138 && $-0.471$ \\
&                  LDA$^b$ & 16.60 & 1.141 && $-0.525$ \\
\\
DO$_{23}$ & GGA$^a$ & 17.16 & 1.080 & $\delta_{\textrm{Al}}=+0.0013$ & $-0.502$\\
&&&&$\delta_{\textrm{Zr}}=-0.0239$     \\
& LDA$^b$ & 16.35 & 1.087 & $\delta_{\textrm{Al}}=-0.0021$ & $-0.548$\\
&&&&$\delta_{\textrm{Zr}}=-0.0273$     \\

&Exp.\footnote{Neutron diffraction\cite{AMA95}} & 17.25 & 1.0775 &$\delta_{\textrm{Al}}=+0.0004$  & \\
&&&&$\delta_{\textrm{Zr}}=-0.0272$\\
&Exp.\footnote{Calorimetry\cite{MES93}} &&&& $-0.502 \pm 0.014$ \\
\hline
\end{tabular}\end{center}
\end{minipage}
\end{table}

\subsection{Cluster expansion of the formation energy}
\label{chap_cluster}

In order to express the formation energy of any Al-Zr compound based on a perfect fcc lattice,
we make a cluster expansion\cite{SAN84} of our FP-LMTO calculations to fit a generalized Ising model.
This allows us to obtain the energy of any configuration of the fcc lattice.

Considering a binary alloy of N sites on a rigid lattice,
its configuration can be described through an Ising model
by the vector $\boldsymbol{\sigma} =\{\sigma_1, \sigma_2, \ldots, \sigma_N\}$
where the pseudo-spin configuration variable $\sigma_i$ is equal to
$\pm1$ if an A or B atom occupies the site $i$.
Any structure is then defined by its density matrix $\rho^s$,
$\rho^s(\boldsymbol{\sigma})$ being the probability of finding the
structure $s$ in the configuration $\boldsymbol{\sigma}$.

With any cluster of $n$ lattice points $\alpha=\{i_1,i_2,\ldots,i_n\}$
we associate the multisite correlation function
\begin{equation}
\zeta_{\alpha}^s
= \trace  \rho^s \prod_{i\in\alpha}\sigma_i
= \frac{1}{2^N} \sum_{\boldsymbol{\sigma}} \rho^s(\boldsymbol{\sigma}) \prod_{i\in\alpha}\sigma_i
\label{correlation},
\end{equation}
where the sum has to be performed over the $2^N$ possible configurations of
the lattice.

Clusters related by a translation or a symmetry operation of the point
group of the structure have the same correlation functions.
Denoting by $D_{\alpha}$ the number of such equivalent clusters per
lattice site, or degeneracy, the energy, like any other configurational function,
can be expanded in the form \cite{SAN84}
\begin{equation}
E = \sum_{\alpha} D_{\alpha} J_{\alpha} \zeta_{\alpha}^s
\label{ce},
\end{equation}
where the sum has to be performed over all non equivalent clusters and
the cluster interaction $J_{\alpha}$ is independent of the structure.

The cluster expansion as defined by equation~\ref{ce} cannot be
used directly: a truncated approximation of the sum has to be
used. The truncation is made with respect to the number of points
contained in a cluster, thus assuming that order effects on energy
are limited to a small set of lattice points. It is truncated too
with respect to distance between sites. Long range interactions
are important mostly if one wants to fully reproduce elastic
effects \cite{LAK92}.

\begin{table}[!hbt]
\begin{minipage}{\textwidth}
\renewcommand{\footnoterule}{}
\caption{Formation energy relative to pure fcc elements for Al-Zr
compounds lying on a perfect fcc lattice ($a=a_{Al}=4.044$~\AA)
obtained from a direct FP-LMTO calculations and from its cluster
expansion.} \label{energy}
\begin{center}
\begin{tabular}{lcc.{-1}.{-1}}
\hline
& Pearson & Structure &\multicolumn{2}{c}{$E^{form}$ (eV/atom)} \\
\cline{4-5}
& symbol & type & \multicolumn{1}{c}{FP-LMTO} &  \multicolumn{1}{c}{CE}  \\
\hline
Al (fcc)                              & cF4  & Cu           &  0.      &  0.    \\
Al$_4$Zr (D1$_{\textrm{a}}$) & tI10 & MoNi$_4$ & -0.421 & -0.491\\
Al$_3$Zr (L1$_2$)                & cP4 & Cu$_3$Au & -0.728 & -0.671 \\
Al$_3$Zr (DO$_{22}$)         & tI8   & Al$_3$Ti   & -0.617 & -0.643 \\
Al$_3$Zr (DO$_{23}$)         & tI16 & Al$_3$Zr   & -0.690 & -0.657 \\
Al$_2$Zr ($\beta$)              & tI6   & MoSi$_2$  & -0.513 & -0.482 \\
AlZr (L1$_0$)                     & tP4  & AuCu        &-0.803  & -0.780 \\
AlZr (L1$_1$)                     & hR32 & CuPt       & -0.448 & -0.466 \\
AlZr (CH40)                       & tI8    & NbP         &-0.643  & -0.723 \\
AlZr (D4)                           & cF32 &
?\footnote{Description of structures D4 and Z2 can be found in Ref. \cite{LU91}}
&-0.489 & -0.414 \\
AlZr (Z2)                            & tP8   & ?$^a$            &-0.345 & -0.333 \\
Zr$_2$Al ($\beta$)              & tI6   & MoSi$_2$  & -0.443 & -0.482 \\
Zr$_3$Al (L1$_2$)              & cP4 & Cu$_3$Au & -0.640 & -0.603 \\
Zr$_3$Al (DO$_{22}$)         & tI8   & Al$_3$Ti   & -0.570 & -0.574 \\
Zr$_3$Al (DO$_{23}$)         & tI16 & Al$_3$Zr   & -0.603 & -0.589 \\
Zr$_4$Al (D1$_{\textrm{a}}$) & tI10 & MoNi$_4$ & -0.390 & -0.437 \\
Zr (fcc)                              & cF4  & Cu           &   0. & 0. \\
\hline
\end{tabular}
\end{center}
\end{minipage}
\end{table}

\begin{table}[!hbt]
\caption{Cluster expansion of the formation energy.}
\label{coef_ce}
\begin{center}
\begin{tabular}{c.{-1}.{-1}}
\hline
&&\multicolumn{1}{c}{$J_\alpha$}  \\
\raisebox{1.5ex}[0pt]{Cluster}  &
\multicolumn{1}{c}{\raisebox{1.5ex}[0pt]{$D_{\alpha}$}} &
\multicolumn{1}{c}{(eV/atom)}  \\
\hline
\{0\}   & 1     & -4.853\\
\{1\}   & 1     & 0.933    \\
\{2,1\} & 6     & 97.5\times10^{-3}  \\
\{2,2\} & 3     & -28.4\times10^{-3} \\
\{3,1\} & 8     &  4.2\times10^{-3} \\
\{4,1\} & 2     &  13.1\times10^{-3} \\
\hline
\end{tabular}
\end{center}
\end{table}

We use in the expansion of the energy six different clusters: the
empty cluster \{0\}, the point cluster \{1\}, the pairs of first
and second nearest neighbors \{2,1\} and \{2,2\}, the triangle of
first nearest neighbors \{3,1\}, and the tetrahedron of   first
nearest neighbors \{4,1\}. The corresponding cluster interactions
are obtained by making a least square fit of compound energies
calculated with FP-LMTO. All 17 used compounds are lying on a
perfect fcc lattice: energies are calculated without any
relaxation of the volume, of the shape of the unit cell, or of the
atomic positions. The lattice parameter used is the one which minimizes
the cohesive energy of pure Al, $a=a_{Al}=4.044$~\AA. We choose to
fit the cluster expansion for the equilibrium lattice parameter of
Al because we are interested in describing thermodynamics of the
Al rich solid solution as well as of Al$_3$Zr precipitates in the
L1$_2$ structure. These precipitates have an equilibrium lattice
parameter close to the one of pure Al, $a=4.073$~\AA{ }as obtained
from FP-LMTO calculations with GGA, and during the nucleation
stage they are coherent with the Al matrix. Consequently, such an
expansion should be able to give a reasonable thermodynamic
description of the different configurations reached during this
precipitation stage where precipitates are coherent.

Coefficients of the cluster expansion of the energy are given in table \ref{coef_ce}.
Comparing the values of the many-body interactions, we see that the main contribution
to the energy arises from the pair interactions and that the 3- and 4-point cluster
contributions are only corrections.
Signs of pair interactions reflect the tendency of Al and Zr atoms
to form heteroatomic first nearest neighbor pairs
and homoatomic second nearest neighbor pairs.

In table \ref{energy}, we compare the formation energies of the
different compounds directly obtained from FP-LMTO calculations
with the ones given by their cluster expansion. The standard
deviation equals 41~meV/atom and the maximal difference is
79~meV/atom. This could have been improved by including more
clusters in the expansion of the energy or by using a mixed-space
cluster expansion\cite{LAK92}. Nevertheless, this would not have
changed the main characteristics of the Al-Zr system, \ie the
short range order tendency given by pair interactions, as well as
the dependence on local environment of the interactions given by
3- and 4-point cluster interactions. In order to be able to build
a realistic kinetic model and to run Monte Carlo simulations in a
reasonable amount of time, we have to keep the thermodynamic
description of Al-Zr system as simple as it can be. Therefore we
do not try to improve expansion convergence and we focus our work
on the influence of the 3- and 4-point cluster interactions on
the thermodynamic and kinetic properties.

\subsection{Phase diagram}
\label{phase_diag_chap}

\begin{figure}[!hbp]
\begin{center}
\includegraphics[width=0.7\textwidth]{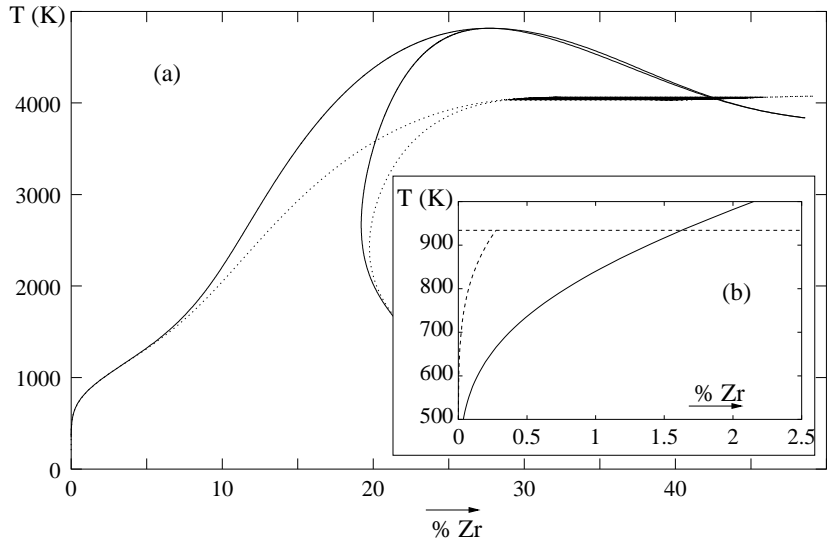}
\end{center}
\figcaption{Al rich part of the phase diagram corresponding to the
equilibrium between the fcc solid solution and the L1$_2$
structure given by our set of parameters (table \ref{coef_ce}).
(a)~Comparison of the phase diagrams obtained with pair,
triangle, and tetrahedron interactions (solid line) and the one
obtained with only pair interactions (dotted line).
(b)~Comparison with the predicted metastable solubility
limit\cite{CLO02} (dashed line).} \label{phase_diag}
\end{figure}

We use the cluster-variation method (CVM) \cite{KIK51} in the
tetrahedron-octahedron (TO) approximation \cite{SAN80, MOH85} to
study the equilibrium between the fcc Al-rich solid solution and
the L1$_2$ structure (Fig. \ref{phase_diag}) corresponding to
energy parameters of table \ref{coef_ce}. At low temperature, the
2 sublattices of the L1$_2$ structure remain highly ordered, as at
the experimental peritectic melting temperature ($T\sim934$~K) the
Zr concentrations of the two sublattices are respectively 100. and 1.8~at.\%.
Turning out the energy coefficients of the first nearest neighbor
triangle and tetrahedron ($J_3=J_4=0$), we see that these
many-body interactions have a thermodynamic influence only at high
temperature (Fig.~\ref{phase_diag}~(a)), as for temperatures below
1000~K the phase diagram remains unchanged with or without these
interactions.

In figure \ref{phase_diag}~(b), we compare the Zr solubility limit in the fcc solid solution
corresponding to the present work energy parameters
with our previous estimation of this metastable solubility limit \cite{CLO02}.
We should point out that the solubility limit obtained with the parameters given by table
\ref{coef_ce} corresponds to a coherent equilibrium between the fcc solid solution
and the L1$_2$ structure as the energy coefficients of the expansion have been
calculated for a perfect fcc lattice at the parameter of pure Al.
This leads to a destabilization of the ordered phase and this is the main reason
why we obtain a higher solubility than the estimated one corresponding to
the equilibrium between incoherent phases.
Another reason is that we use the cluster expansion to compute Al$_3$Zr cohesive energy,
and thus get a small error on this energy,
whereas in our previous study we directly used the value given by FP-LMTO calculation.

\section{KINETIC MODEL}
\label{kinetic_chap}

In order to be able to build an atomistic kinetic model, we have to generalize our thermodynamic description
of the Al-Zr  binary system to the one of the Al-Zr-Vacancy ternary system.
To do so, we recast first the spin-like formalism of the cluster expansion  into the more convenient one
of the lattice gas formulation using occupation numbers\cite{DUCASTELLE}.
This will allow us to obtain effective interactions for the different configurations of
the tetrahedron of first nearest neighbors and of the pair of second nearest neighbors.
Atom-vacancy interactions can then be introduced quite easily.

\subsection{Effective interactions}
\label{appendix_Ising}

Instead of using the pseudo-spin variables $\sigma_n$ as we did in
chap \ref{chap_cluster}, this will be easier for the following to
work with occupation numbers $p^{i}_{n}$, $p^{i}_{n}$ being equal
to 1 if an atom of type $i$ occupies the site $n$ and to 0
otherwise. In a binary alloy, occupation numbers and pseudo-spin
variable at site $n$ are related by
\begin{equation}
p^{A}_{n}  =  \frac{1+\sigma_n}{2}, \qquad\textrm{and}\qquad
p^{B}_{n}  =  \frac{1-\sigma_n}{2} .
\end{equation}

For the Al-Zr binary system, we included in our truncated cluster
expansion of the energy first nearest neighbor interactions up to
the pair, triangle, and tetrahedron clusters and a second nearest
neighbor pair interaction. Thus, using the occupation numbers
$p^{i}_{n}$, the expression of the energy becomes
\begin{equation}
E = \frac{1}{4 N_s} \sum_{\substack{ n,m,p,q \\ i,j,k,l}} \epsilon_{ijkl}^{(1)} p^{i}_{n} p^{j}_{m}  p^{k}_{p}  p^{l}_{q}
+ \frac{1}{2 N_s} \sum_{\substack{ r,s \\ i,j}} \epsilon_{ij}^{(2)} p^{i}_{r} p^{j}_{s} ,
\label{short_Ising}
\end{equation}
where the first sum runs over all sites $(n,m,p,q)$ forming a
first nearest neighbor tetrahedron
 and all their different configurations $(i,j,k,l)$,
 and the second sum over all sites $(r,s)$ forming a second nearest neighbor pair
 and all their different configurations $(i,j)$.
 $N_s$ is the number of lattice sites,
$\epsilon_{ijkl}^{(1)}$ the effective energy of a first nearest
neighbor tetrahedron in the configuration $(i,j,k,l)$, and
$\epsilon_{ij}^{(2)}$ the effective energy of a second nearest
neighbor pair in the configuration $(i,j)$.

Writing the energy with these effective interactions increases the
number of dependent variables. Therefore several choices of these
effective energies correspond to the same cluster expansion, then
to the same thermodynamic and kinetic properties. If we make the
assumption that second nearest neighbor interactions do not
contribute to the cohesive energy of pure elements, \ie
$\epsilon_{AA}^{(2)}=0$ and   $\epsilon_{BB}^{(2)}=0$, we obtain
as many effective interactions as parameters in the truncated
cluster expansion. Such an assumption does not have any physical
influence and it just guarantees that homo-atomic effective
interactions, $\epsilon_{AAAA}^{(1)}$ and $\epsilon_{AA}^{(2)}$,
do not depend on the on the nature of B atom. Effective energies
of the first nearest neighbor tetrahedron in its different
configurations are then related to the cluster expansion
coefficients by the equations
\begin{equation}
\left(\begin{array}{c}
\epsilon_{AAAA}^{(1)}  \\ \epsilon_{AAAB}^{(1)} \\ \epsilon_{AABB} ^{(1)}  \\  \epsilon_{ABBB}^{(1)} \\ \epsilon_{BBBB}^{(1)}
\end{array}\right)
= \frac{1}{12} \left(\begin{array}{rrrrr}
 6 &  6 &  6 &   6  & 6 \\
 6 &  3 &  0 & -3 &-6 \\
6 & 0 & -2 & 0 & 6 \\
6 & -3 & 0 & 3 & -6 \\
6 & -6 & 6 & -6 & 6
\end{array}\right)
\left(\begin{array}{c}
D_0 J_0 + D_{2,2} J_{2,2} \\ D_1 J_1  \\ D_{2,1} J_{2,1} \\  D_{3,1} J_{3,1} \\  D_{4,1} J_{4,1}
\end{array}\right),
\label{CE2short_Ising}
\end{equation}
and the second nearest neighbor pair interaction by the equation
\begin{equation}
\epsilon_{AB}^{(2)}   =   -\frac{2}{3}  D_{2,2} J_{2,2} .
\end{equation}
For Al-Zr binary system, tetrahedron effective interactions
corresponding to the cluster expansion of chap.~\ref{chap_cluster}
can be found in table \ref{effective_energies_table}: two sets are
given depending if $J_3$ and $J_4$ are taken from the cluster
expansion of table \ref{coef_ce} or are supposed equal to zero.
For both sets $\epsilon_{AB}^{(2)} = +0.057$~eV.

\subsection{Decomposition of effective interactions}

As we wrote before, several sets of effective interactions produce
the same cluster expansion. In the following, we generate the set
of interactions useful for our kinetic model for which we have to
count bonds we break for vacancy-atom exchange.

Different contributions are included in the effective energy
$\epsilon_{ijkl}^{(1)}$. One part of the energy is due to the
bonding corresponding to the six different pairs of atoms
contained in the tetrahedron, each of these pairs belonging to two
different tetrahedrons. Then one has to add corrections due to
order on the four triangles contained in the tetrahedron and
another correction due to order on the tetrahedron itself. This
decomposition leads to the relation
\begin{eqnarray}
\epsilon_{ijkl}^{(1)} &=& \frac{1}{2} \left( \epsilon_{ij}^{(1)} + \epsilon_{ik}^{(1)} + \epsilon_{il}^{(1)}
+ \epsilon_{jk}^{(1)} + \epsilon_{jl}^{(1)} + \epsilon_{kl}^{(1)} \right)  \nonumber \\
&&+ \left(  \widetilde{\epsilon}_{ijk}^{(1)} + \widetilde{\epsilon}_{ijl}^{(1)}
+\widetilde{\epsilon}_{ikl}^{(1)} +\widetilde{\epsilon}_{jkl}^{(1)} \right)
+ \widetilde{\epsilon}_{ijkl}^{(1)} ,
\label{short2long_Ising}
\end{eqnarray}
where $\epsilon_{ij}^{(1)}$ is the effective energy of the first
nearest neighbor pair in the configuration $(i,j)$ and
$\widetilde{\epsilon}_{ijk}^{(1)}$ and
$\widetilde{\epsilon}_{ijkl}^{(1)}$ the corrections to add to pair
energy due to order on triangles and on the tetrahedron

Using the previous breakdown of the tetrahedron effective energy,
the expression \ref{short_Ising} of the energy becomes
 \begin{eqnarray}
E &=& \frac{1}{2 N_s} \sum_{\substack{ n,m \\ i,j}} \epsilon_{ij}^{(1)} p^{i}_{n} p^{j}_{m}
+  \frac{1}{3 N_s} \sum_{\substack{ n,m,p \\ i,j,k}} \widetilde{\epsilon}_{ijk}^{(1)} p^{i}_{n} p^{j}_{m}  p^{k}_{p} \nonumber \\
&& + \frac{1}{4 N_s} \sum_{\substack{ n,m,p,q \\ i,j,k,l}} \widetilde{\epsilon}_{ijkl}^{(1)} p^{i}_{n} p^{j}_{m}  p^{k}_{p}  p^{l}_{q}
+ \frac{1}{2 N_s} \sum_{\substack{ r,s \\ i,j}} \epsilon_{ij}^{(2)} p^{i}_{r} p^{j}_{s} .
\label{long_Ising}
\end{eqnarray}
As this is just another mathematical way to rewrite the cluster expansion \ref{ce} of the energy,
the following relations holds :
\begin{subeqnarray}
D_0 J_0 + D_{2,2} J_{2,2} & = & \frac{3}{2} \left( \eAA + 2\eAB +\eBB \right) \slabel{CE_empty}\\
&& + \eAAA + 3\eAAB + 3\eABB + \eBBB \nonumber \\
&& + \frac{1}{8} \left( \eAAAA + 4\eAAAB + 6\eAABB + 4\eABBB + \eBBBB \right)
 \nonumber \\
D_1 J_1 & = & 3 \left( \eAA - \eBB \right) \slabel{CE_point}\\
&& + 3 \left( \eAAA + \eAAB - \eABB - \eBBB \right) \nonumber \\
&& + \frac{1}{2} \left( \eAAAA + 2\eAAAB   - 2\eABBB - \eBBBB \right)
 \nonumber \\
D_{2,1} J_{2,1} & = & \frac{3}{2} \left( \eAA - 2\eAB +\eBB \right) \slabel{CE_pair}\\
&& + 3 \left( \eAAA - \eAAB - \eABB + \eBBB \right) \nonumber \\
&& + \frac{3}{4} \left( \eAAAA - 2\eAABB  + \eBBBB \right)
 \nonumber \\
D_{3,1} J_{3,1} & = &  \eAAA - 3\eAAB + 3\eABB - \eBBB \slabel{CE_tri} \\
&& + \frac{1}{2} \left( \eAAAA - 2\eAAAB   + 2\eABBB - \eBBBB \right)
\nonumber \\
D_{4,1} J_{4,1} & = & \frac{1}{8} \left( \eAAAA - 4\eAAAB + 6\eAABB - 4\eABBB + \eBBBB \right)
\slabel{CE_tetra}
\end{subeqnarray}

As we want $\widetilde{\epsilon}_{ijk}^{(1)}$ to be the energetic
corrections due to order on triangles, the contribution to
$J_{2,1}$ of $\eAAA$, $\eAAB$, \ldots{ } must equal zero (second
term in right hand side of eq. \ref{CE_pair}). For the same
reason, the contribution to $J_{2,1}$ and $J_{3,1}$ of $\eAAAA$,
$\eAAAB$, \ldots{ } must equal zero (last term in right hand side
of eq. \ref{CE_pair} and \ref{CE_tri}). We require too that
triangle and tetrahedron order corrections do not contribute to
the cohesive energy of pure elements, as we did for second nearest
neighbor pair interactions. Thus, $\eAAA=\eBBB=0$ and
$\eAAAA=\eBBBB=0$. With these restrictions, all parameters
entering in the expression \ref{long_Ising} of the energy are well
determined.

The first nearest neighbor pair effective energies are thus
\begin{subeqnarray}
\eAA & = & \frac{1}{6} \left(
D_0 J_0 + D_1 J_1 + D_{2,1} J_{2,1}  + D_{2,2} J_{2,2} + D_{3,1} J_{3,1} + D_{4,1} J_{4,1} \right) \\
\eAB & = &   \frac{1}{6} \left( D_0 J_0 - D_{2,1} J_{2,1}  + D_{2,2} J_{2,2} + D_{4,1} J_{4,1} \right) \\
\eBB & = &   \frac{1}{6} \left(
D_0 J_0 - D_1 J_1 + D_{2,1} J_{2,1}  + D_{2,2} J_{2,2} - D_{3,1} J_{3,1} + D_{4,1} J_{4,1} \right) ,
\end{subeqnarray}
the order corrections on first nearest neighbor triangle
\begin{equation}
\eAAB = - \eABB = - \frac{1}{6} D_{3,1} J_{3,1} ,
\end{equation}
and the order corrections on first nearest neighbor tetrahedron
\begin{subeqnarray}
&\eAAAB = \eABBB = - D_{4,1} J_{4,1} &\\
&\eAABB  =  0 .&
\end{subeqnarray}
Inverting the system \ref{CE2short_Ising}, one can easily express all these quantities from the effective tetrahedron
energies $\epsilon_{ijkl}$ too.

\subsection{Interactions with vacancy}

Within the previous formalism, we can easily introduce
atom-vacancy interactions. These interactions are a simple way to
take into account the electronic relaxations around the vacancy.
Without them, the vacancy formation energy $E^{for}_V$ in a pure
metal would necessarily equal the cohesive energy
($E^{for}_V=0.69$~eV \cite{LANDOLT25} and $E^{coh}=3.36$~eV for
fcc Al).

We only consider first-nearest neighbor interactions with
vacancies and we do not include any order correction on triangle
and tetrahedron configurations containing at least one vacancy,
\ie
$\widetilde{\epsilon}^{(1)}_{ijV}=\widetilde{\epsilon}^{(1)}_{ijkV}=0$
where $i$, $j$, and $k$ are any of the species Al, Zr, and V. The
vacancy formation energy in a pure metal A is then given by
\begin{equation}
E^{for}_V = 8\epsilon_{AAAV}^{(1)} - 6 \epsilon_{AAAA}^{(1)}
= 12 \epsilon_{AV}^{(1)} - 6 \epsilon_{AA}^{(1)}
\end{equation}

The interaction $\epsilon_{AlV}^{(1)}$ is deduced from the experimental value of
the vacancy formation energy in pure Al.
For the interaction $\epsilon_{ZrV}^{(1)}$, we assume that the vacancy formation energy
in the fcc structure is the same as in the hcp one, these two structures being quite similar.
The only experimental information we have concerning this energy is
$E^{for}_V > 1.5$~eV~\cite{LANDOLT25}.
We thus use the ab-initio value calculated by Le Bacq \etal \cite{LEB99}, $E^{for}_V = 2.07$~eV.
This value is calculated at the equilibrium volume of Zr and cannot be used directly
to obtain $\epsilon_{ZrV}^{(1)}$ as this interaction should correspond in our model
to the equilibrium lattice parameter of pure Al.
We have to use instead the vacancy formation enthalpy
\begin{equation}
H^{for}_V = E^{for}_V  + P \delta \Omega^{for}_V ,
\end{equation}
where $\delta \Omega^{for}_V = -1.164$~\AA$^3$ is the vacancy
formation volume in pure Zr \cite{LANDOLT25}, and $P$ is the pressure
to impose to Zr to obtain a lattice parameter equal to the one of
Al. $P$ is calculated from the bulk modulus $B=91$~GPa of fcc Zr
and the equilibrium volumes of Al and Zr, $\Omega^0_{Al}=16.53$
~\AA$^3$ and  $\Omega^0_{Zr}=23.36$ ~\AA$^3$, these three
quantities being obtained from our FP-LMTO calculations. This
gives us the value  $H^{for}_V = 1.88$~eV for the vacancy
formation enthalpy in pure Zr at the lattice parameter of pure Al.

We use the experimental value of the divacancy binding energy
$E^{bin}_{2V}=0.2$~eV \cite{LANDOLT25} in order to compute a vacancy-vacancy interaction,
$\epsilon_{VV}^{(1)} = E^{bin}_{2V} - \epsilon_{AlAl}^{(1)} + 2 \epsilon_{AlV}^{(1)}$.
If we do not include this interaction and set it equal to zero instead,
we obtain the wrong sign for the divacancy binding energy,
divacancies being thus more stable than two monovacancies.
This does not affect our Monte Carlo simulations as we only include one vacancy in the simulation box,
but this will have an influence if we want to build a mean field approximation of our diffusion model.

We thus managed to add vacancy contributions to our thermodynamic
description of the Al-Zr binary system. Using the breakdown
\ref{short2long_Ising} of the first nearest neighbor tetrahedron
interaction, we can obtain the effective energies corresponding to
the 15 different configurations a tetrahedron can have in the
ternary system. These effective energies are presented in table
\ref{effective_energies_table} for the cases where energy
correction due to order on first nearest neighbor triangle and
tetrahedron are assumed different from zero or not ($J_3$ and
$J_4$ given by table \ref{coef_ce} or $J_3=J_4=0$).

\begin{table}[!hbt]
\caption{Effective energies of the first nearest neighbor
tetrahedron for Al-Zr-V ternary system. The set with order
correction corresponds to the values $J_3$ and $J_4$ given by the
cluster expansion of table \ref{coef_ce} and the set without order
correction assumes $J_3=J_4=0$.} \label{effective_energies_table}
\begin{center}
\begin{tabular}{c.{-1}.{-1}}
\hline
Configuration & \multicolumn{2}{c}{Effective energy (eV)} \\
\cline{2-3}
 & \multicolumn{1}{c}{with order} &
 \multicolumn{1}{c}{without order} \\
& \multicolumn{1}{c}{correction} &
 \multicolumn{1}{c}{correction} \\
\hline
Al Al Al Al & -1.680 & -1.680\\
Al Al Al Zr & -2.257 & -2.214\\
Al Al Zr Zr & -2.554 & -2.554\\
Al Zr Zr Zr & -2.707 & -2.698\\
Zr Zr Zr Zr & -2.647 & -2.647\\
Al Al Al V & -1.174  & -1.174\\
Al Al Zr V & -1.567  & -1.561\\
Al Zr Zr V & -1.748  & -1.754\\
Zr Zr Zr V & -1.751  & -1.751\\
Al Al V V & -0.518   & -0.518\\
Al Zr V V & -0.758   & -0.758\\
Zr Zr V V & -0.804   & -0.804\\
Al V V V & +0.288    & +0.288\\
Zr V V V & +0.194    & +0.194\\
V V V V & +1.243     & +1.243\\
\hline
\end{tabular}
\end{center}
\end{table}

With this set of thermodynamic parameters, we calculate the binding energy between a Zr solute atom
and a vacancy in Al,
\begin{equation}
E^{bin}_{ZrV} = 2 \left( \epsilon_{AlAlAlAl}^{(1)} + \epsilon_{AlAlZrV}^{(1)}
- \epsilon_{AlAlAlZr}^{(1)} - \epsilon_{AlAlAlV}^{(1)} \right) .
\end{equation}
The value obtained, $E^{bin}_{ZrV}=+0.369$~eV, agrees with the experimental observation that
there is no attraction between  Zr solute atoms and vacancies in Al \cite{LANDOLT25, SIM77}.

\subsection{Migration barriers}

Diffusion occurs via vacancy jumps towards one of its twelve first
nearest neighbors. The vacancy exchange frequency with a neighbor
of type $A$ ($A=$Al or Zr) is given by
\begin{equation}
\Gamma_{A-V}=\nu_A \exp{\left(-\frac{E^{act}_A}{k_BT} \right)} ,
\end{equation}
where $\nu_A$ is an attempt frequency and the activation energy
$E^{act}_A$ is the energy change required to move the $A$ atom
from its initial stable position to the saddle point position. It
is computed as the difference between the contribution $e^{sp}_A$
of the jumping atom to the saddle point energy and the
contributions of the vacancy and of the jumping atom to the
initial energy of the stable position. This last contribution is
obtained by considering all bonds which are broken by the jump,
\ie all pair interactions the vacancy and the jumping atoms are
forming as well as all order corrections on triangles and
tetrahedrons containing the jumping atom,
\begin{equation}
E^{act}_A = e^{sp}_A - \sum_j \epsilon_{Vj}^{(1)} - \sum_{j\ne
V}\epsilon_{Aj}^{(1)} - \sum_{jk}\widetilde{\epsilon}_{Ajk}^{(1)}
-\sum_{jkl}\widetilde{\epsilon}_{Ajkl}^{(1)}
-\sum_j\epsilon_{Aj}^{(2)}.
\end{equation}

The attempt frequency $\nu_A$ and the contribution $e^{sp}_A$of
the jumping atom to the saddle point energy can depend on the
configuration \cite{LEB02}. Nevertheless, we do not have enough
information to see if such a dependence holds in the case of Al-Zr
alloys. We thus assume that these parameters depend only on the
nature of the jumping atom, which gives us four purely kinetic
parameters to adjust.

The contribution of Al to the saddle point energy, $e^{sp}_{Al}$,
is deduced from the experimental value of the vacancy migration
energy in pure Al, $E^{mig}_V=0.61$~eV \cite{LANDOLT25}, and the
attempt frequency $\nu_{Al}$ from the experimental Al
self-diffusion coefficient, $D_{Al^*}=D_0 \exp{(-Q/k_BT)}$, the
self-diffusion activation energy $Q$ being the sum of the vacancy
formation and migration energies in pure Al and
$D_0=1.73\times10^{-5}$~m$^2$.s$^{-1}$ \cite{LANDOLT26}.

To calculate $\nu_{Zr}$ and $e^{sp}_{Zr}$, we use the experimental
value\footnote{Diffusion coefficient measured in the temperature
range between 800 and 910~K.} of the diffusion coefficient of Zr
impurity in Al, $D_{Zr^*} = 728\times10^{-4}
\exp{(-2.51\textrm{~eV}/k_BT)}$~m$^{2}$.s$^{-1}$ \cite{LANDOLT26,
MAR73}. The kinetic parameters can be deduced from this
experimental data by using the five frequency model for solute
diffusion in fcc lattices \cite{BOC96}, if we make the assumption
that there is no correlation effect. We check afterwards that such
an assumption is valid: at $T=500$~K the correlation factor is
$f_{Zr^*}=1$ and at $T=1000$~K $f_{Zr^*}=0.875$. Correlation
effects are thus becoming more important at higher temperature but
they can be neglected in the range of temperature used in the
fitting procedure.

\begin{table}[!hbp]
\caption{Kinetic parameters for a thermodynamic description of
Al-Zr binary with and without energy corrections due to order on
first nearest neighbor triangle and tetrahedron.}
\label{kinetic_para}
\begin{center}
\begin{tabular}{l.{-1}.{-1}}
\hline 
& \multicolumn{1}{c}{with order} & \multicolumn{1}{c}{without order} \\
& \multicolumn{1}{c}{correction} & \multicolumn{1}{c}{correction} \\
\hline
$e^{sp}_{Al}$ & -8.219\textrm{~eV} &  -8.219\textrm{~eV} \\
$e^{sp}_{Zr}$ & -11.286\textrm{~eV} & -10.942\textrm{~eV} \\
$\nu_{Al}$     & 1.36\times 10^{14}\textrm{~Hz} & 1.36\times 10^{14}\textrm{~Hz} \\
$\nu_{Zr}$     & 4.48\times 10^{17}\textrm{~Hz} & 4.48\times 10^{17}\textrm{~Hz} \\
\hline
\end{tabular}
\end{center}
\end{table}

So as to study the influence on kinetics of energy corrections due to order on
triangles and tetrahedrons, we fit another set of kinetic parameters corresponding
to a thermodynamic description of Al-Zr binary with only pair interactions
(\ie $J_3=J_4=0$, or equivalently
$\widetilde{\epsilon}_{ijk}^{(1)}=\widetilde{\epsilon}_{ijkl}^{(1)}=0$).
This other set of kinetic parameters presented in table \ref{kinetic_para}
reproduces as well coefficients for Al self-diffusion and for Zr impurity diffusion,
the only difference being that these kinetic parameters
correspond to a simpler thermodynamic description of Al-Zr binary.


\section{DIFFUSION IN SOLID SOLUTION} \label{diffusion_chap}

Diffusion in Al-Zr solid solutions can be fully characterized by
the tracer correlation coefficients $f_{Al}$ and $f_{Zr}$
and by the phenomenological Onsager coefficients $L_{AlAl}$, $L_{AlZr}$, and $L_{ZrZr}$.
These coefficients link fluxes of diffusing species, $\mathbf{J}_{Al}$ and $\mathbf{J}_{Zr}$,
to their chemical potential gradients \cite{PHILIBERT, ALLNATT} through the relations
\begin{eqnarray}
\mathbf{J}_{Al} & = & -L_{AlAl} \mathbf{\nabla} \mu'_{Al} / k_BT
- L_{AlZr} \mathbf{\nabla} \mu'_{Zr} / k_BT \nonumber \\
\mathbf{J}_{Zr} & = & -L_{AlZr}  \mathbf{\nabla} \mu'_{Al} / k_BT
 -L_{ZrZr} \mathbf{\nabla} \mu'_{Zr} / k_BT .
\end{eqnarray}
Chemical potentials entering these equations are relative to the
vacancy chemical potential, $\mu'_{Al}=\mu_{Al}-\mu_V$ and
$\mu'_{Zr}=\mu_{Zr}-\mu_V$. We use to express diffusion fluxes the
Onsager reciprocity condition, $L_{AlZr}=L_{ZrAl}$.

These coefficients can be used in finite-difference diffusion code
so as to study "industrial" processes where diffusion is involved
(precipitation, solidification, homogenization, \ldots)
\cite{BOR00}. One way to obtain these coefficients is to adapt
Calphad methodology to kinetics, \ie to guess an expression for
$L_{AB}$ describing its variation with temperature and composition
of the alloy and to adjust the model parameters on a large kinetic
database \cite{AND92,CAM02}. On the other hand one can use an
atomistic model as the one described in chap.~\ref{kinetic_chap}
to obtain the phenomenological coefficients \cite{MAR90, DES97,
DES98}. If one carefully applies the same mean field approximation
for thermodynamics and kinetics it is possible to get the whole
Onsager matrix and not only diagonal terms and to catch all
correlation effects \cite{NAS00}. Such an approach compared to the
previous one does not need a huge experimental database. Moreover,
as it is based on a realistic description of diffusion at the
atomic scale, it appears safer to extrapolate kinetic quantities
out of the range (composition or temperature) used in the fitting
procedure.

In this study, we do not use any mean -field approximation to
calculate phenomenological coefficients, but obtain them directly
from kinetic Monte Carlo simulations by using generalization of
the Einstein formula for tracer diffusion due to Allnatt
\cite{ALL82, ALLNATT}
\begin{equation}
L_{AB} = \dfrac{ \left< \mathbf{\Delta R}_A \cdot \mathbf{\Delta R}_{B} \right> }{6 \Delta t},\qquad A, B = Al, Zr,
\label{Onsager_eq}
\end{equation}
where the brackets indicate a thermodynamic ensemble average
and $\mathbf{\Delta R}_A$ is the sum of total displacement $\mathbf{\Delta r}_i$
of all atoms $i$ of type $A$ during time $\Delta t$,
\begin{equation}
\mathbf{\Delta R}_A = \sum_{i \in A} \mathbf{\Delta r}_i .
\end{equation}

We use residence time algorithm to run kinetic Monte Carlo calculations.
The simulation box contains 125000 lattice sites,
one of this site being occupied by a vacancy.
Sum of total displacements $\mathbf{\Delta R}_{Al}$ and $\mathbf{\Delta R}_{Zr}$
in equation \ref{Onsager_eq} are computed
for a time interval corresponding to $\sim10^6$ vacancy jumps,
and their thermodynamic averages are obtained through
simulations of $10^9$ vacancy jumps.
Such a big number of jumps is necessary to converge thermodynamic averages
entering in the calculation of $L_{AlZr}$ and $L_{ZrZr}$,
whereas $L_{AlAl}$ converges more quickly.
This is due to the difference of diffusion coefficients between Al and Zr.

\begin{figure}[!hbt]
\begin{center}
\includegraphics[width=0.7\textwidth]{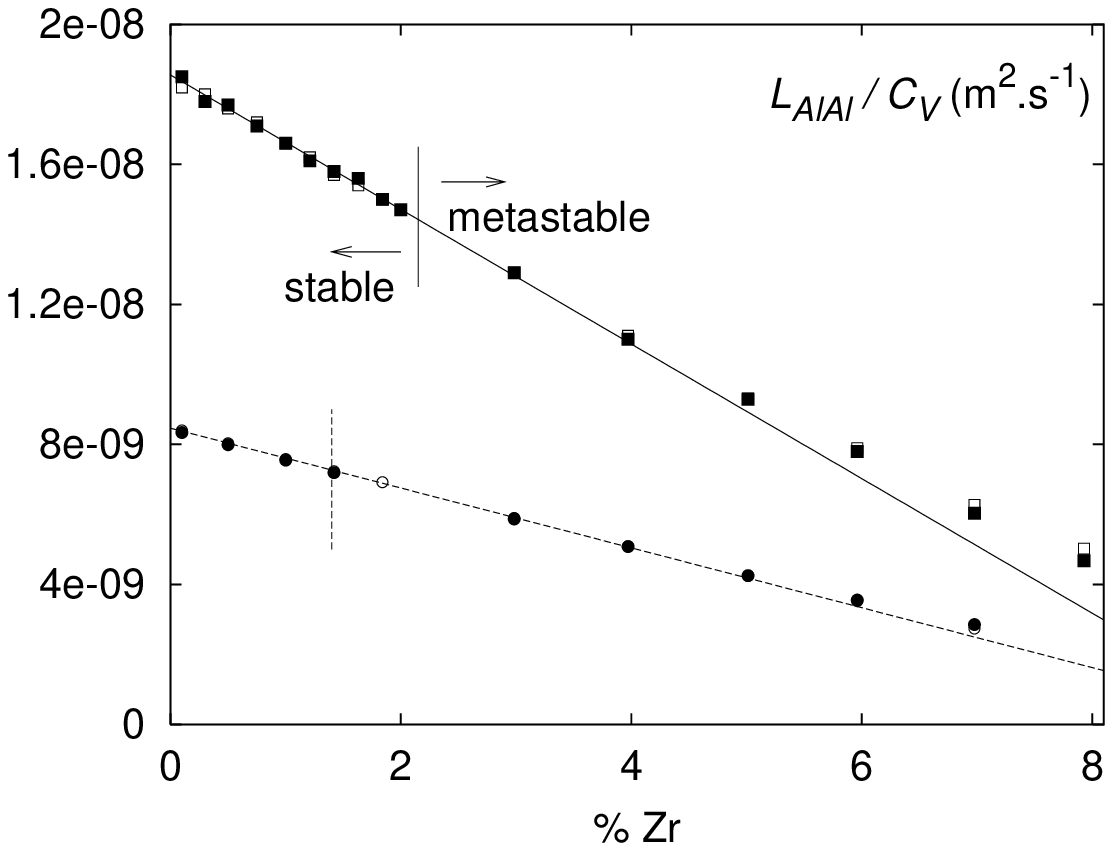}
\hfill
\includegraphics[width=0.7\textwidth]{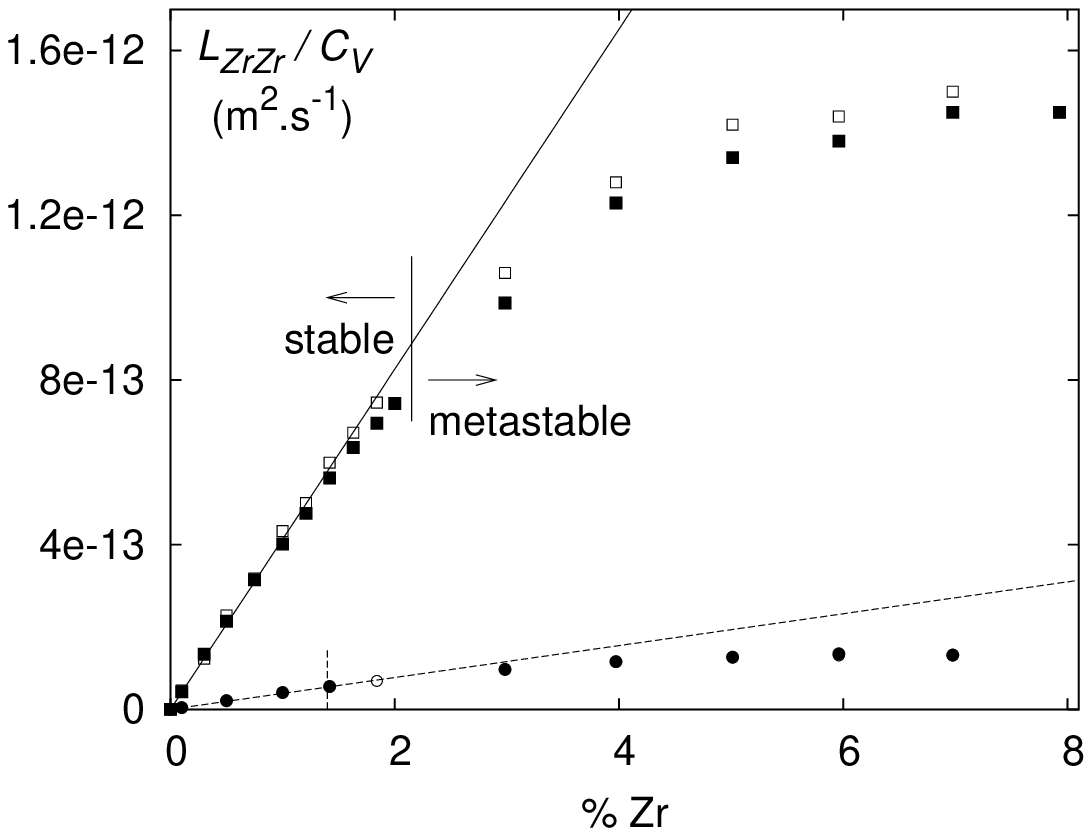}
\end{center}
\figcaption{Onsager coefficients $L_{AlAl}$ and $L_{ZrZr}$.
Squares and solid lines correspond to $T=1000$~K and circles and
dashed lines to $T=900$~K. The vertical lines indicate the
corresponding solubility limit obtained from CVM calculations.
Full symbols correspond to the set of parameters with order
corrections on triangles and tetrahedrons and open symbols to the
set without order corrections.} \label{Onsager_fig}
\end{figure}

Results of calculations are presented in figure \ref{Onsager_fig}
for two different temperatures, $T=1000$~K and $T=900$~K,
and different Zr concentration from 0 to 8~at.\%.
For the off-diagonal coefficient $L_{AlZr}$ of Onsager matrix,
dispersion is too important to get a precise value of thermodynamic average\footnote{
$L_{AlZr} = 0 \pm 10^{-12}$~m$^{2}$.s$^{-1}$ at $T=1000$~K
and $L_{AlZr} = 0 \pm 10^{-13}$~m$^{2}$.s$^{-1}$ at $T=900$~K}.
We interpret this as an indication that this coefficient can be neglected
in this range of temperature and concentration.

Onsager coefficients are calculated for Zr concentration
corresponding to the stable as well as to the metastable solid
solution, the limit being given by the CVM calculations of
chap.~\ref{chap_cluster}. For calculations in the metastable solid
solution, thermodynamic averages are computed during the
incubation stage of precipitation kinetics when no stable
precipitate is present in the simulation box
(chap.~\ref{chap_preci}). $L_{AlAl}$ behavior deviates only
slightly from its linear extrapolation from the stable solid
solution, but for $L_{ZrZr}$ it seems that no extrapolation from
the stable to the metastable solid solution is possible.

So as to see the influence of triangles and tetrahedrons interactions,
we ran simulations with only pair interactions considering the corresponding kinetic
parameters of table~\ref{kinetic_para}.
One can directly see on figure \ref{Onsager_fig} that these two sets of parameters
reproduce the same experimental data, \ie the self-diffusion coefficient
\begin{equation}
D_{Al^*} = f_0 \lim_{C_{Zr} \rightarrow 0} L_{AlAl} ,
\end{equation}
where $f_0=0.78145$ for a fcc lattice,
and the Zr impurity diffusion coefficient
\begin{equation}
D_{Zr^*} = \lim_{C_{Zr} \rightarrow 0} L_{ZrZr} / C_{Zr} .
\end{equation}

Order corrections mainly affect $L_{ZrZr}$. This coefficient is
slightly lower when one considers energy corrections due to order
on triangles and tetrahedrons. The difference increases with Zr
concentration and thus in the metastable solid solution: these
order corrections lead to a slight slowdown of Zr diffusion. The
two thermodynamic models are equivalent at these temperatures (\cf
phase diagram on Fig.~\ref{phase_diag}~(a)). As a consequence
kinetic behaviors obtained from them are really close.

\begin{figure}[!hbt]
\begin{center}
\includegraphics[width=0.49\textwidth]{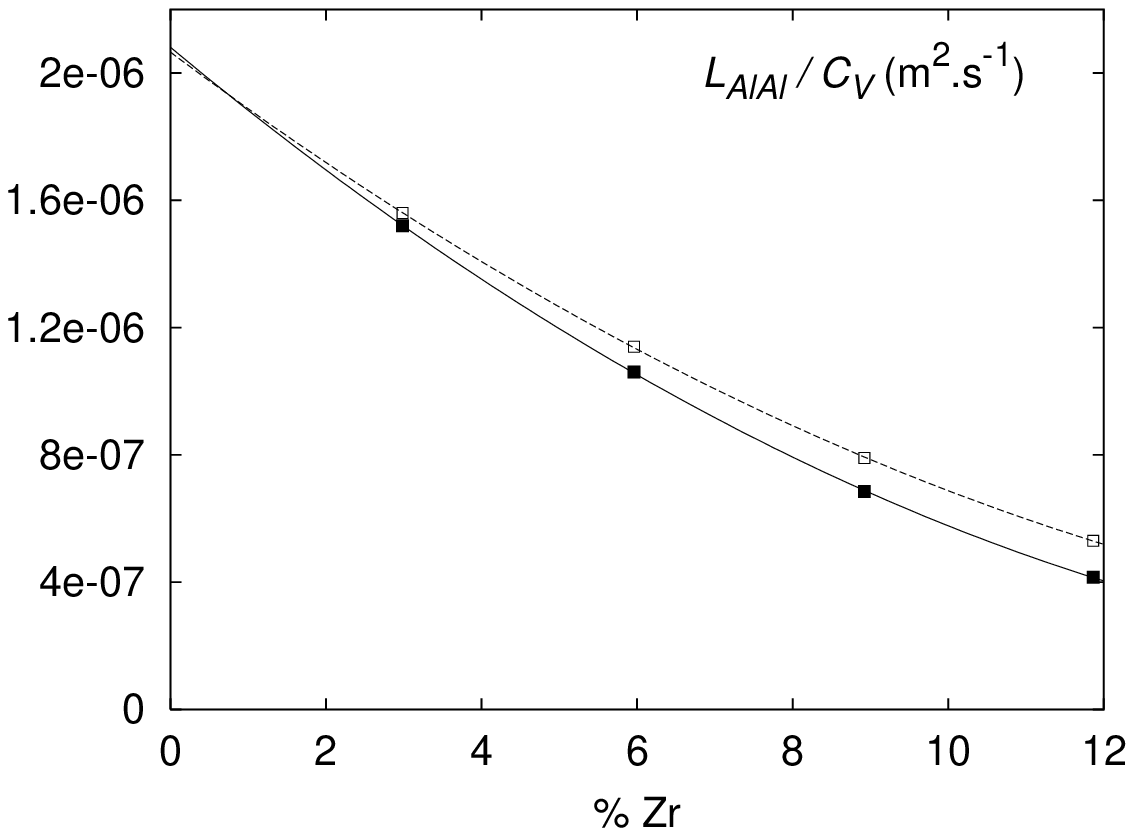}
\includegraphics[width=0.49\textwidth]{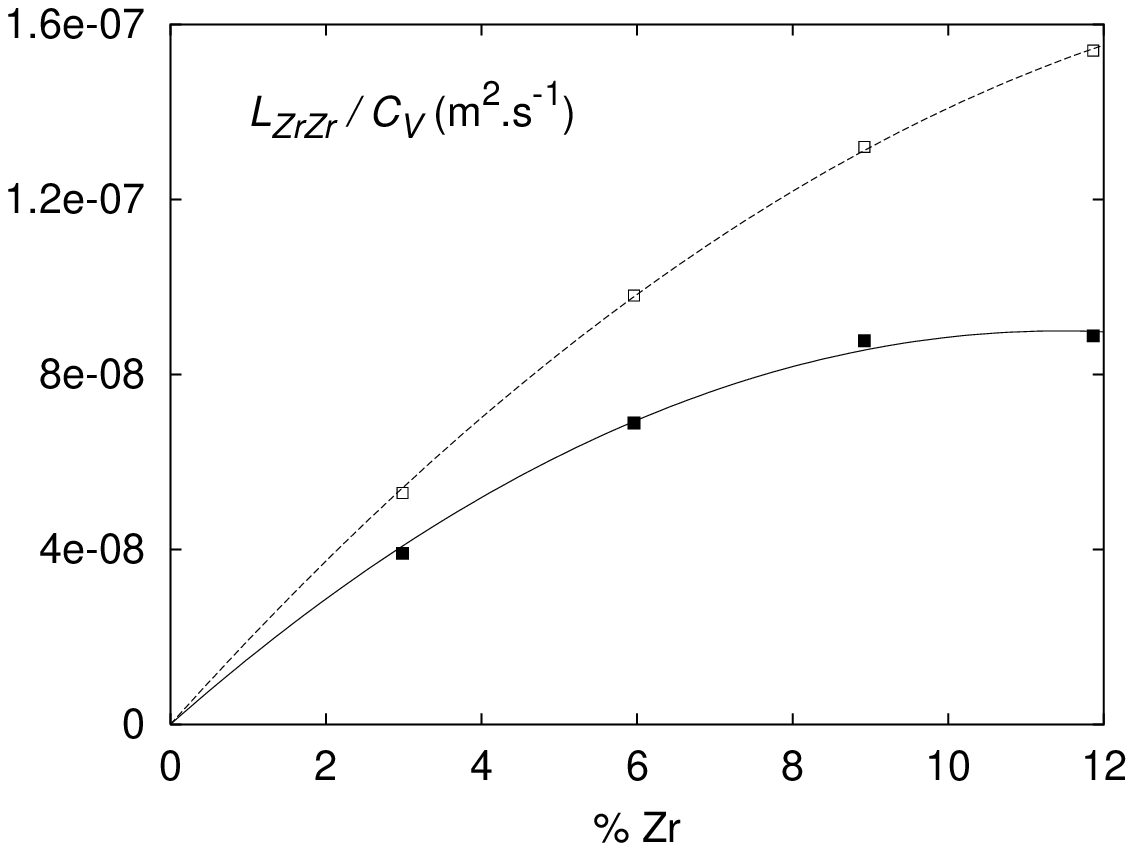}
\includegraphics[width=0.49\textwidth]{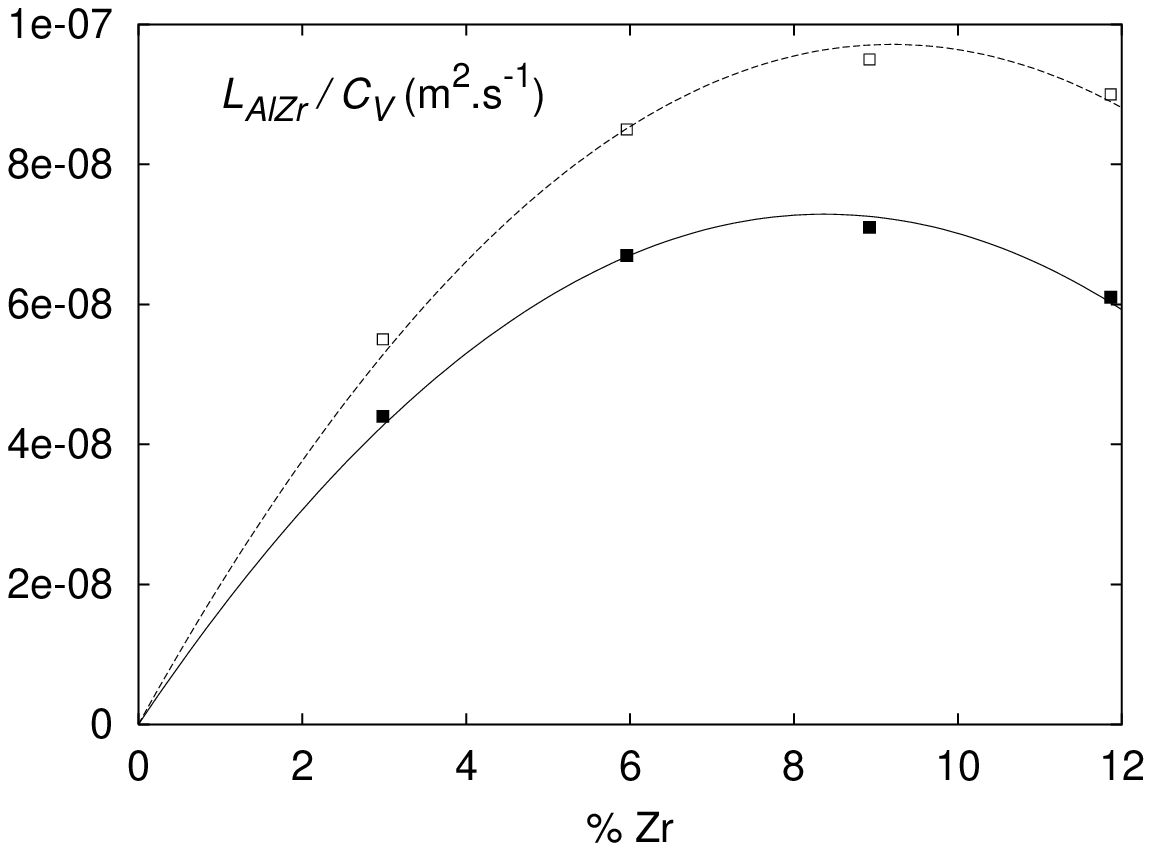}
\end{center}
\figcaption{Onsager coefficients $L_{AlAl}$, $L_{ZrZr}$, and
$L_{AlZr}$ calculated at $T=3000$~K. Full symbols and solid lines
correspond to the set of parameters with order corrections on
triangles and tetrahedrons and open symbols and dashed lines to
the set without order corrections.} \label{Onsager_highT_fig}
\end{figure}

At higher temperatures, triangle and tetrahedron interactions
change the phase diagram (Fig.~\ref{phase_diag}~(a)). This
thermodynamic influence leads to a kinetic change too: at
$T=3000$~K, Onsager coefficients are lower when considering these
multisite interactions (Fig. \ref{Onsager_highT_fig}). One should
notice that correlation effects cannot be neglected at this
temperature as $L_{AlZr}$ is far from being null. Thus one is not
allowed anymore to assume Onsager matrix as diagonal. With
triangle and tetrahedron interactions, Al$_3$Zr precipitate is
more stable, which means that order effects are stronger. The
kinetic corrolary of this thermodynamic influence is that they
slow down diffusion.

\section{KINETICS OF PRECIPITATION} \label{chap_preci}

Precipitation kinetics have been obtained by Monte Carlo simulations
for four different supersaturations of the solid solution
($C^0_{Zr}=5$, 6, 7, and 8~at.\%) at $T=1000$~K.
At this temperature, the equilibrium concentration is $C^{eq}_{Zr}=2.1$~at.\%.
The simulation box contains 125000 lattice sites and
its starting configuration is a completely disordered (random) solid solution.

\subsection{Short range order parameters}

\begin{figure}[!hbt]
\begin{center}
\includegraphics[width=0.48\textwidth]{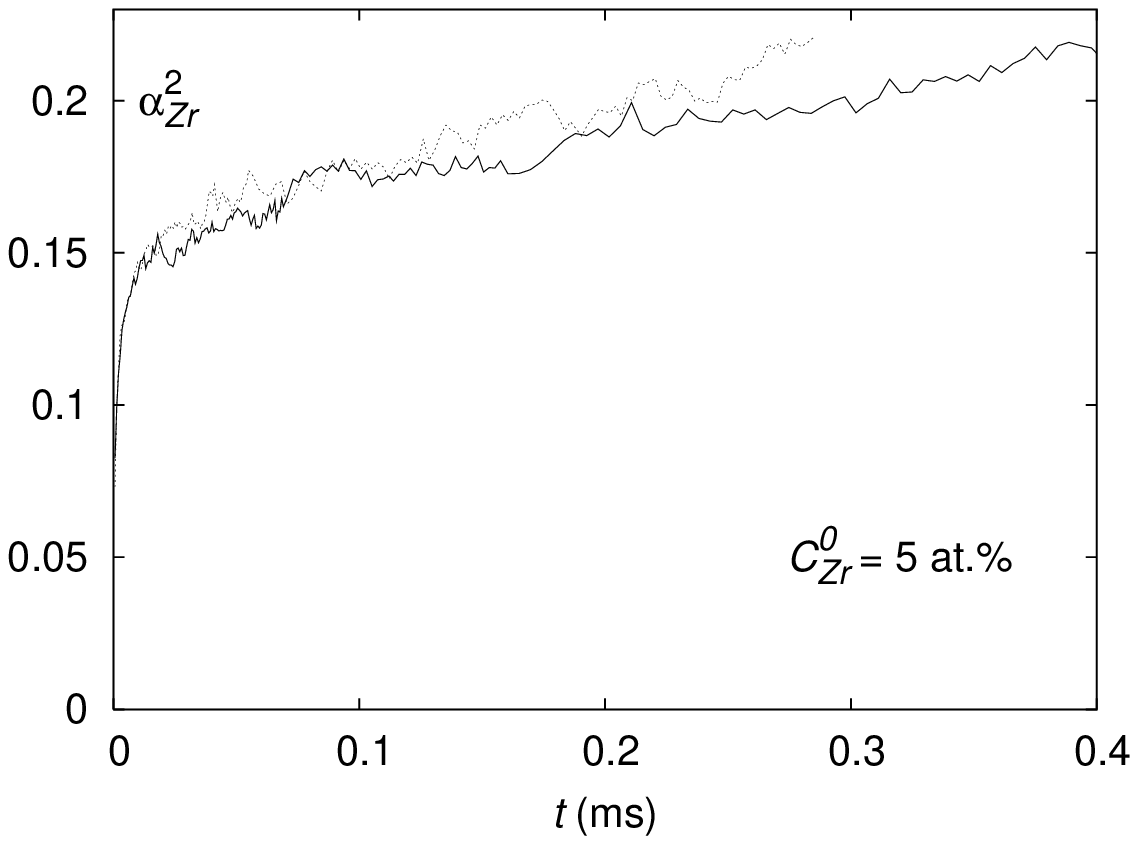}
\hfill
\includegraphics[width=0.48\textwidth]{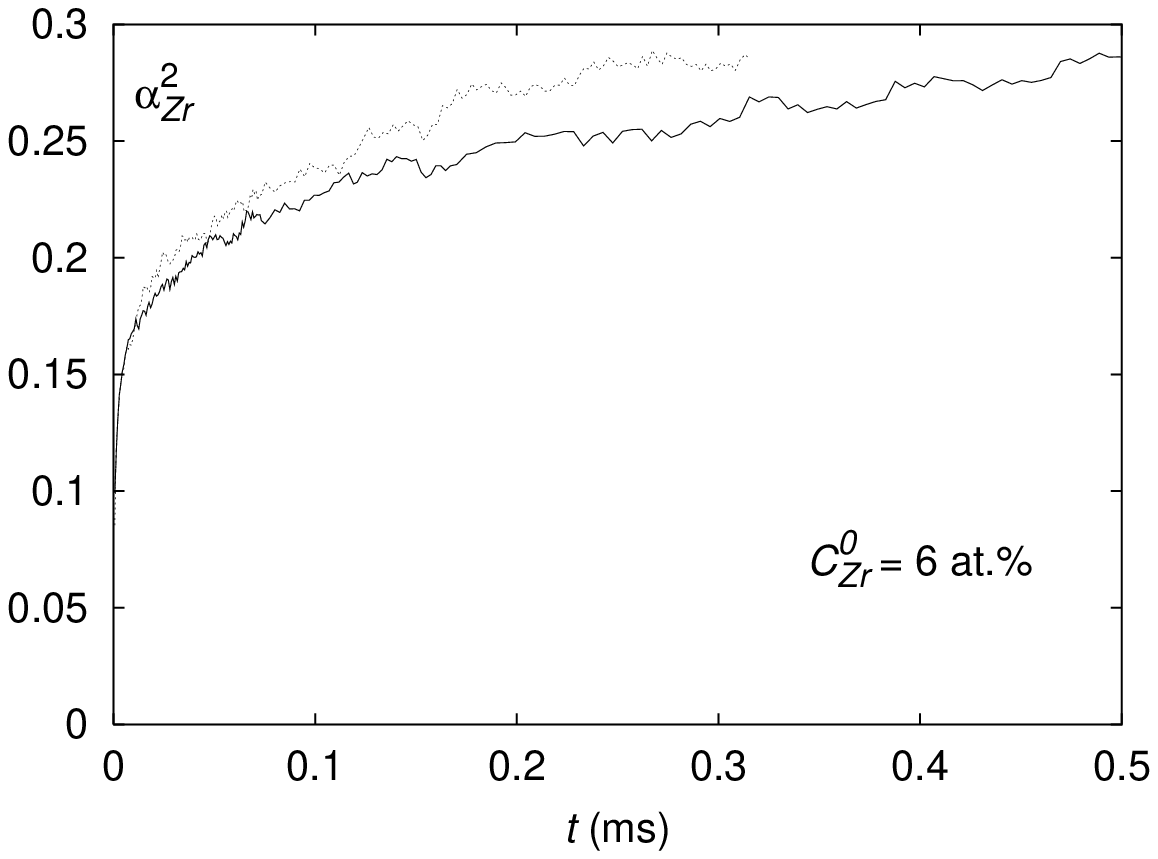}
\includegraphics[width=0.48\textwidth]{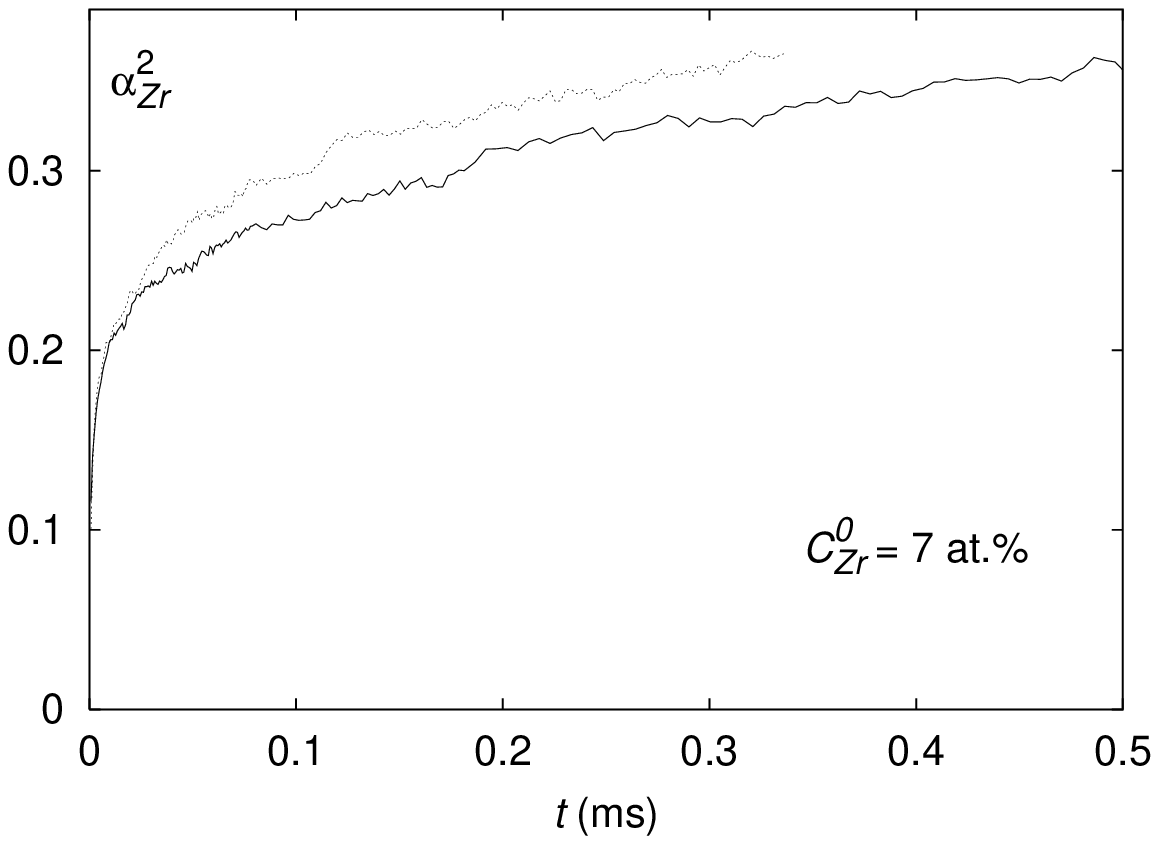}
\hfill
\includegraphics[width=0.48\textwidth]{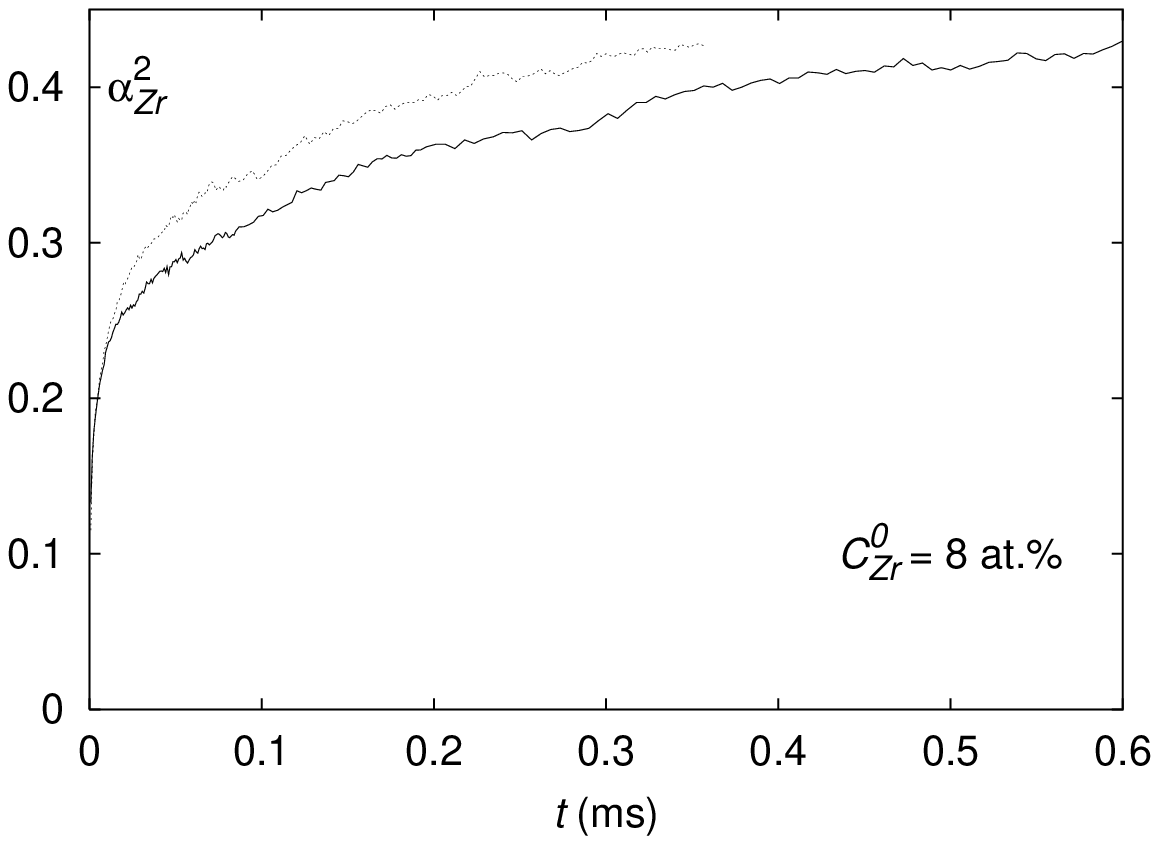}
\end{center}
\figcaption{Evolution of second nearest neighbor short range order
of Zr atoms, $\alpha^2_{Zr}$, at $T=1000$~K and four different
nominal concentrations $C^0_{Zr}$. Full and dotted lines are
respectively for the set of parameters with and without order
corrections on triangles and tetrahedrons.} \label{sro_fig}
\end{figure}

The quantities of interest to follow the global evolution of
precipitation during the simulation are Warren-Cowley short
range order (SRO) parameters\cite{DUCASTELLE}. SRO parameters for
first-nearest neighbors evolve too quickly to give any really
significant information on precipitation state. During
simulation first steps,  Zr atoms surround themselves with Al.
Once this local equilibrium for first nearest neighborhood is
reached, the corresponding SRO parameters do not evolve anymore.
On the other hand, SRO parameters for second nearest neighbors
slowly evolve until the end of the simulation. For Zr atoms, it is
defined as
\begin{equation}
\alpha^2_{Zr} = \frac{ \left< p_n^{Zr} \right>_{Zr, 2} - C^0_{Zr} }{1-C^0_{Zr}},
\end{equation}
where $\left< p_n^{Zr} \right>_{Zr, 2} $ stands for the average of
occupation numbers $p_n^{Zr}$ on all second nearest neighbors of
Zr atoms. For a randomly distributed configuration of the alloy
(initial configuration) $\alpha^2_{Zr}=0$, whereas for the L1$_2$
structure $\alpha^2_{Zr}=1$. Looking at fig.~\ref{sro_fig}, one
sees that $\alpha^2_{Zr}$ evolves more quickly with the set of
parameters with only pair interactions than with triangle and
tetrahedron interactions. At first glance, this is in agreement
with the slight difference on $L_{ZrZr}$ measured in the
metastable solid solution at this temperature (Fig.
\ref{Onsager_fig}) for the two set of parameters. So as to see if
the difference of precipitation kinetics can be understood only in
terms of a difference of diffusion speed or is due to another
factor, we measure the nucleation rate in our simulations and
interpret it with classical theory of nucleation \cite{MAR78,
SOI00, PORTER}.

\subsection{Precipitate critical size}

We first need to give us a criterion to decide which atoms are
belonging to L1$_2$ precipitates. As stable precipitates are
almost perfectly stoichiometric at $T=1000$~K (chap.
\ref{phase_diag_chap}), we only look at Zr atoms and consider for
each Zr atom in L1$_2$ precipitate that three Al atoms are
belonging to the same precipitate. Zr atoms are counted as
belonging to L1$_2$ precipitates if all their twelve first nearest
neighbors are Al atoms and at least half of their six nearest
neighbors are Zr atoms. Moreover, we impose that at least one Zr
atom in a precipitate has its six second nearest neighbors being
Zr, \ie has a first and second nearest neighborhood in perfect
agreement with the L1$_2$ structure.

Classical theory of nucleation predicts there is a critical radius,
or equivalently a critical number $i^*$ of atoms,
below which precipitates are unstable and will re-disolve into the solid solution
and above which precipitates will grow.
$i^*$ is obtained by considering the competition between the interface free energy $\sigma$
and the nucleation free energy per atom $\Delta G^n$,
\begin{equation}
i^* = \frac{2\pi}{3} \left( \frac{a^2\sigma}{\Delta G^n} \right)^3 .
\label{crit_size}
\end{equation}
Clusters of size $i<i^*$ are considered to be local variations of the solid solution composition
and thus are not counted as L1$_2$ precipitates.

\subsection{Nucleation free energy}

\begin{figure}[!hbt]
\begin{center}
\includegraphics[width=0.7\textwidth]{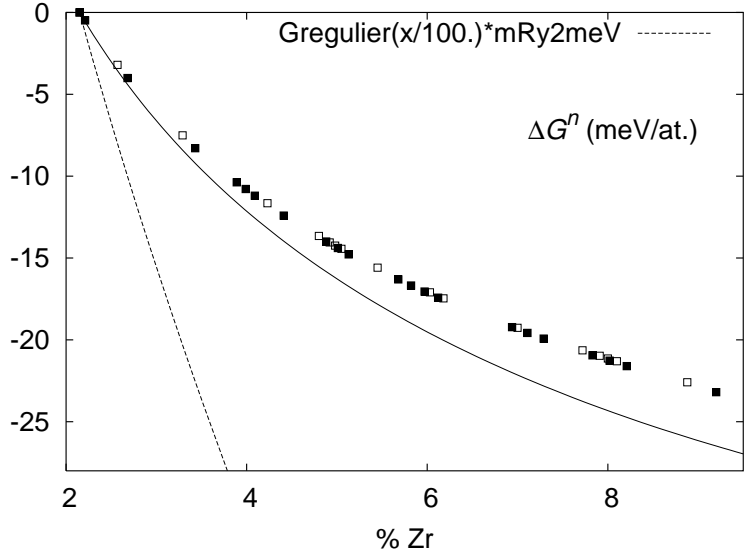}
\end{center}
\figcaption{Nucleation free energy $\Delta G^n$ at $T=1000$~K for
different concentration of the solid solution. Square symbols
correspond to CVM-TO calculation and the line to the ideal
solution approximation (eq. \ref{Gnuc_ideal}). Full and open
symbols are respectively for the set of parameters with and
without order corrections on triangles and tetrahedrons.}
\label{Gnuc_fig}
\end{figure}

The nucleation free energy per atom entering equation \ref{crit_size} is given by \cite{MAR78, PORTER}
\begin{equation}
\Delta G^n = \frac{3}{4} \left( \mu_{Al}(C^{eq}_{Zr}) - \mu_{Al}(C^{0}_{Zr}) \right)
+ \frac{1}{4} \left( \mu_{Zr}(C^{eq}_{Zr}) - \mu_{Zr}(C^{0}_{Zr}) \right) ,
\label{Gnuc}
\end{equation}
where $\mu_{Al}(C_{Zr})$ and $\mu_{Zr}(C_{Zr})$ are the
chemical potentials of respectively Al and Zr components
in the solid solution of concentration $C_{Zr}$,
$C^{eq}_{Zr}$ is the equilibrium concentration of the solid solution,
and $C^{0}_{Zr}$ the nominal concentration.
The factors $3/4$ and $1/4$ arises from the stoichiometry
of the precipitating phase Al$_3$Zr.
Usually the nucleation free energy is approximated by
\begin{equation}
\Delta G^n = \frac{3}{4} k_B T \log{ \frac{1- C^{eq}_{Zr}}{1-C^{0}_{Zr}} }
+ \frac{1}{4} k_B T \log{ \frac{C^{eq}_{Zr}}{C^{0}_{Zr}} }
\label{Gnuc_ideal}
\end{equation}
which is obtained by considering in equation \ref{Gnuc} the
expressions of the chemical potentials for an ideal solution. As
at $T=1000$~K we obtained the same solubility limit,
$C^{eq}_{Zr}=2.1$~at.\%, with or without triangle and tetrahedron
interactions (\cf phase diagram on fig. \ref{phase_diag}), the
approximation \ref{Gnuc_ideal} cannot be used to see if these
interactions have any influence on the nucleation free energy.
Therefore we use CVM-TO to calculate chemical potentials entering
expression \ref{Gnuc}. Looking at figure \ref{Gnuc_fig}, one
should notice that the ideal solution approximation would have
lead to an overestimation of $\Delta G^n$, the error being
$\sim10$\% for the maximal supersaturation considered. With
CVM-TO, we do not obtain any change in the value of the nucleation
free energy depending we are considering or not order corrections
for first nearest neighbor triangle and tetrahedron. Thus slowdown
of precipitation kinetics with these corrections cannot be
explained by a decreasing of the nucleation free energy.

\subsection{Interface free energy}

To determine the precipitate critical size $i^*$ using expression
\ref{crit_size}, we need to know the value of the interface free
energy $\sigma$ too. We calculate this energy at 0~K for different
orientations of the interface. We therefore do not consider any
configurational entropy and simply obtain the interface energy by
counting the number by area unit of wrong "bonds" compared to pure
Al and  Al$_3$Zr in L1$_2$ structure. For (100) and (110)
interfaces there is an ambiguity in calculating such an energy as
two different planes, one pure Al and the other one of
stoichiometry Al$_{1/2}$Zr$_{1/2}$, can be considered as
interface. Considering L1$_2$ precipitates as stoichiometric will
guarantee that to any type of the two possible interfaces is
associated a parallel interface of the other type. Thus for (100)
and (110) interfaces, we consider the average of these two
different interface energies to be meaningful for the parameter
$\sigma$ entering in classical theory of nucleation. For (111)
interface, as only one interface of stoichiometry
Al$_{3/4}$Zr$_{1/4}$ is possible, we do not obtain such an
ambiguity. The energies corresponding to these different
interfaces are
\begin{displaymath}
\sigma_{100} = \frac{1}{\sqrt{2}} \sigma_{110} = \frac{1}{\sqrt{3}} \sigma_{111}
=  \frac{ 2 \epsilon^{(2)}_{AB} - \epsilon^{(2)}_{AA} -  \epsilon^{(2)}_{BB} }{2a^2},
\quad \textrm{with} \quad a^2\sigma_{100}=57.0\textrm{~meV.}
\end{displaymath}
These interface energies only depend on second nearest neighbor
interactions and therefore are the same with or without order
corrections on first nearest neighbor triangle and tetrahedron. To
determine the critical size of precipitates with equation
\ref{crit_size} we use an interface free energy slightly higher
than $\sigma_{100}$, $a^2 \sigma=64.1$~meV. With this interface
free energy, nucleation rate obtained from Monte Carlo simulations
are in better agreement than with $\sigma_{100}$ (Fig.
\ref{nuc_rate_fig}).  As precipitates observed in Monte Carlo
simulations do not exhibit sharp interfaces, this is quite natural
to have to use an energy higher than the minimal calculated one.

\subsection{Nucleation rate}

\begin{figure}[!hbt]
\begin{center}
\includegraphics[width=0.48\textwidth]{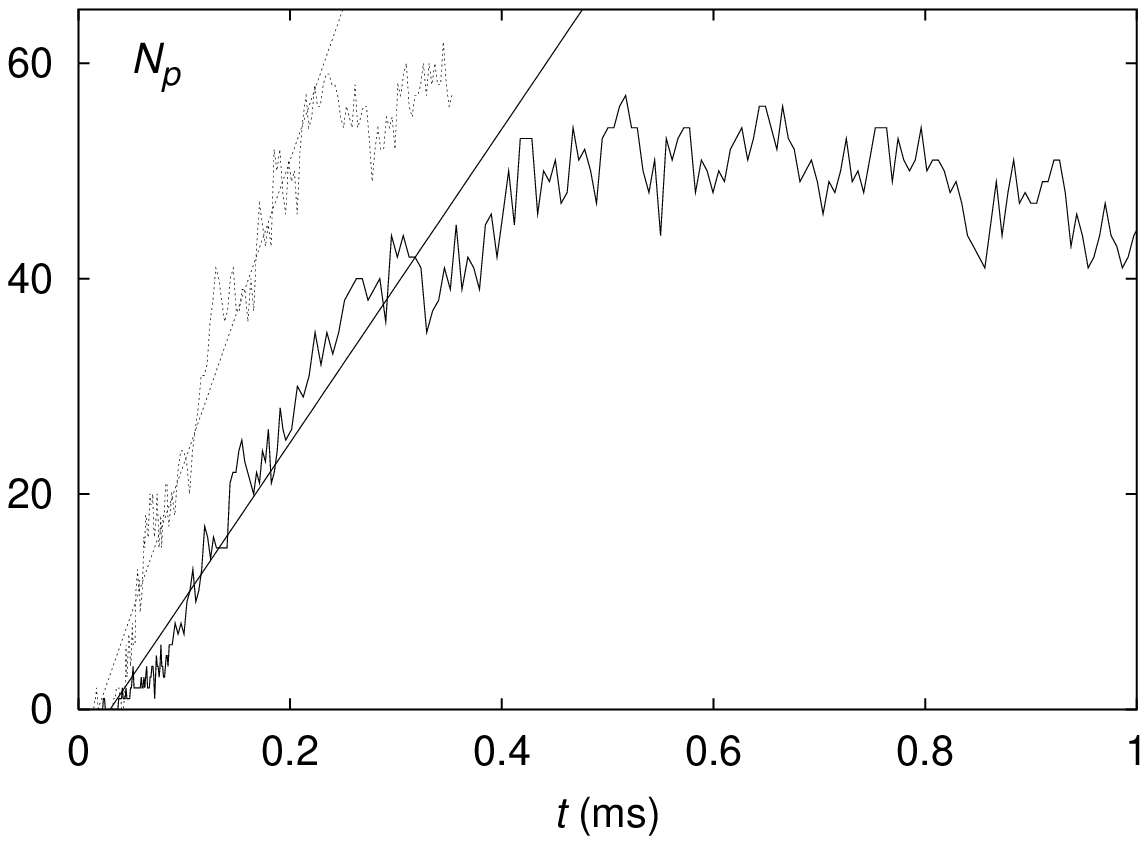}
\hfill
\includegraphics[width=0.48\textwidth]{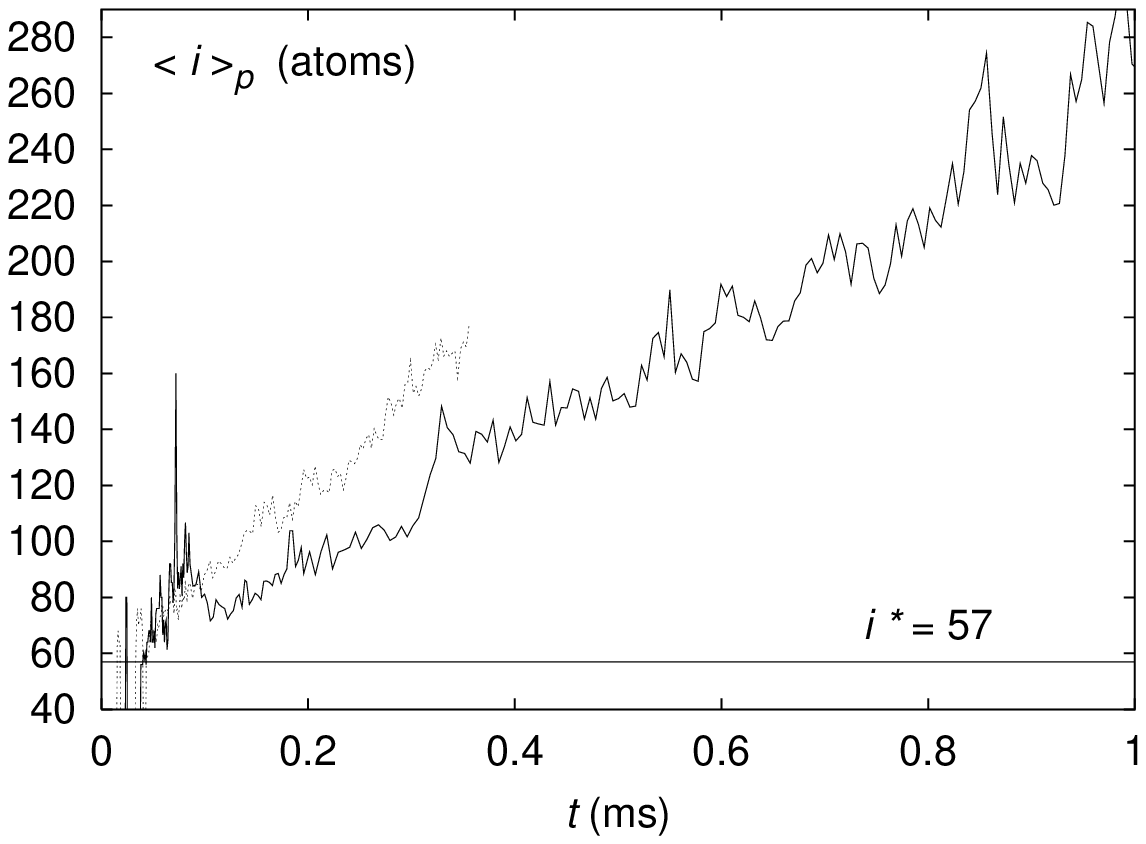}
\includegraphics[width=0.48\textwidth]{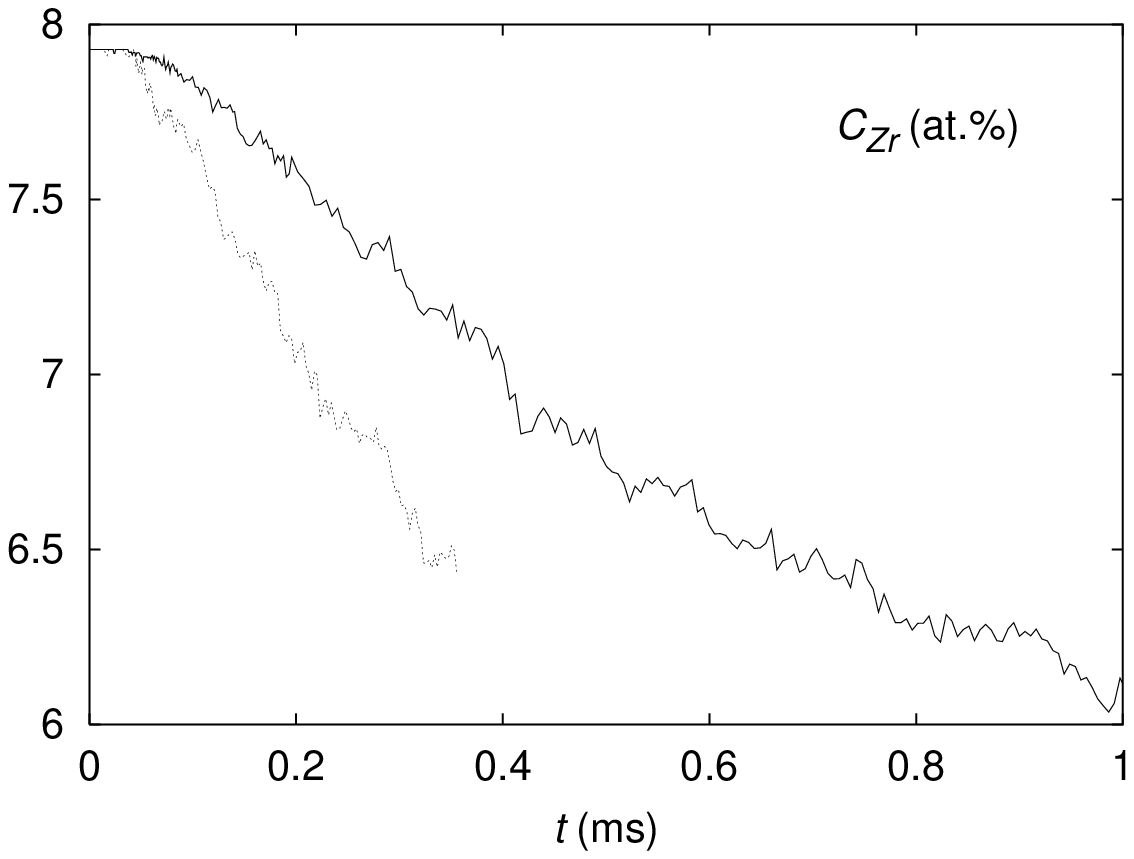}
\end{center}
\figcaption{Kinetics of precipitation for a nominal Zr
concentration $C^0_{Zr}=8$~at.\%: evolution with time of the
number $N_p$ of precipitates in the simulation box, of
precipitates average size $<i>_p$, and of Zr concentration in the
solid solution. Full and dotted lines are respectively for the set
of parameters with and without order corrections on triangles and
tetrahedrons.} \label{kinetic8}
\end{figure}

Critical size for precipitates obtained from these nucleation and
interface free energies are respectively $i^*=187$, 104, 76, and
57 atoms for the different nominal concentrations $C^{0}_{Zr}=5$,
6, 7, and 8~at.\%. We use these critical sizes to determine the
number $N_p$ of supercritical precipitates contained in the
simulation boxes, their average size $\left<i\right>_p$, as well
as the concentration of the solid solution $C_{Zr}$. The variation
with time of these quantities are shown on fig.~\ref{kinetic8} for
the simulation box of nominal concentration $C^0_{Zr}=8$~at.\%.
After an incubation time, one observes a nucleation stage where
the number of precipitates increases linearly until it reaches a
maximum. We then enter into the growth stage: the number of
precipitates does not vary and their size is increasing. At last,
during the coarsening stage, precipitates are still growing but
their number is decreasing. For this concentration, one clearly
sees that precipitation kinetics is faster with only pair
interactions as the number of precipitates is increasing more
rapidly. Moreover precipitates have a bigger size than with
triangle and tetrahedron order corrections.

\begin{figure}[!hbt]
\begin{center}
\includegraphics[width=0.7\textwidth]{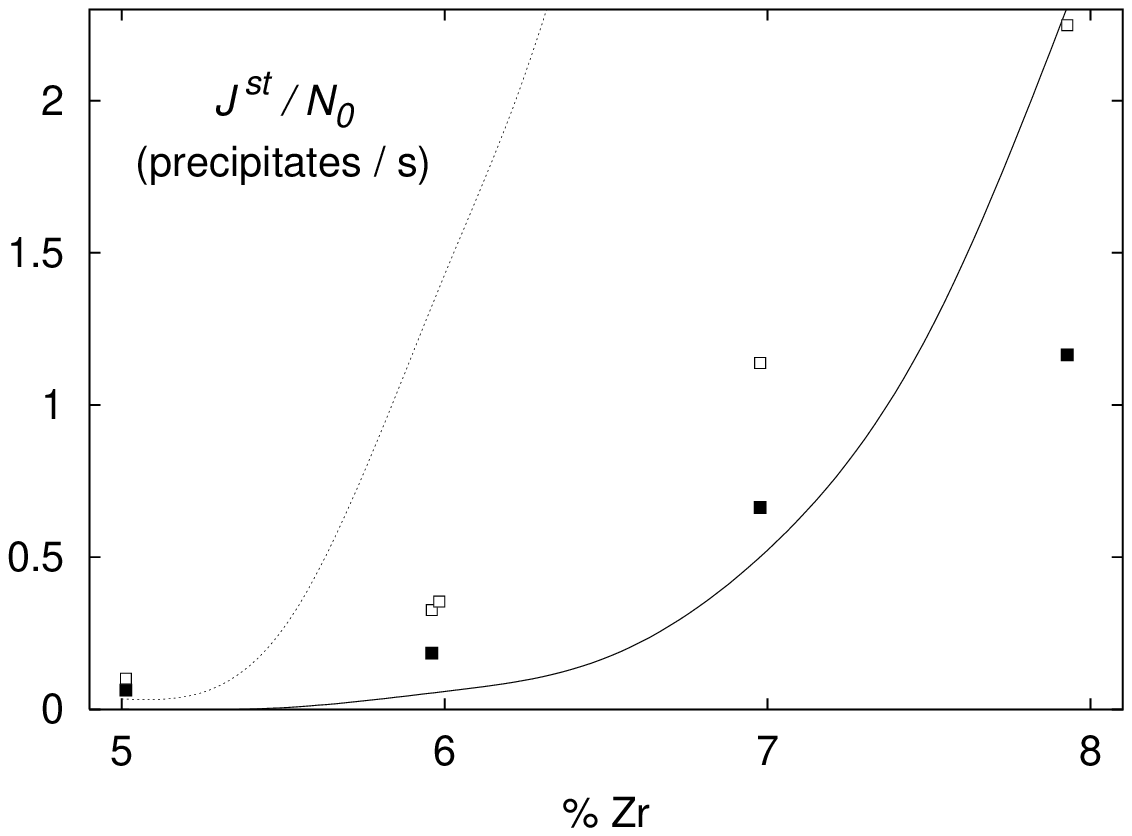}
\end{center}
\figcaption{Evolution of the steady-state nucleation rate $J^{st}$
with the nominal concentration for $T=1000$~K. Full and open
symbols are respectively for the set of parameters with and
without order corrections on triangles and tetrahedrons. The full
line corresponds to the nucleation rate predicted by classical
theory of nucleation with $a^2 \sigma=64.1$~meV and the dotted
line with $a^2 \sigma=a^2\sigma_{100}=57.0$~meV. $J^{st}$ is
normalized by the number of lattice sites in the simulation box,
$N_0=125000$.} \label{nuc_rate_fig}
\end{figure}

The steady-state nucleation rate $J^{st}$ is measured during the
nucleation stage, when the number of precipitates is varying quite
linearly with time.
Slowdown of precipitation kinetics with triangle and tetrahedron order
corrections can be seen on the steady-state nucleation rate (Fig.
\ref{nuc_rate_fig}): without these corrections $J^{st}$ is about
two times higher than when these corrections are included.

In classical theory of nucleation, the steady-state nucleation rate is given
by the expression \cite{MAR78},
\begin{equation}
J^{st} = N_0 Z \beta^* \exp{ -\frac{\Delta G^*}{kT} },
\label{nucleation_rate_eq}
\end{equation}
where $N_0$ is the number of nucleation sites, \ie the number of lattice sites
($N_0=125000$ for Monte Carlo simulations),
$\Delta G^*$ is the nucleation barrier and corresponds to the free energy
of a precipitate of critical size $i^*$,
\begin{equation}
\Delta G^* = \frac{\pi}{3} \frac{ (a^2 \sigma)^3 }{\Delta {G^n}^2},
\end{equation}
$Z$ is the Zeldovitch factor and describes size fluctuations of
precipitates around $i^*$,
\begin{equation}
Z=\frac{1}{2\pi} \frac{\Delta {G^n}^2}{ (a^2 \sigma)^{3/2} \sqrt{kT}},
\end{equation}
and $\beta^*$ is the condensation rate for clusters of critical
size $i^*$. Assuming the limiting step of the adsorption is the
long range diffusion of Zr atoms in the solid solution, the
condensation rate is \cite{MAR78}
\begin{equation}
\beta^* = 8 \pi \frac{a^2 \sigma}{\Delta G^n} \frac{D_{Zr}}{a^2}
C^0_{Zr}. \label{condensation_rate}
\end{equation}
Zr diffusion coefficient is obtained from our measure of Onsager
coefficients in the metastable solid solution (chap.
\ref{diffusion_chap}). Assuming that vacancies are at equilibrium
($\mu_V=0$), its expression is \cite{BOC96, PHILIBERT, ALLNATT}
\begin{equation}
D_{Zr} = \left( L_{ZrZr} - \frac{C^0_{Zr}}{1-C^0_{Zr}}L_{AlZr} \right) \frac{1}{k_B T}
\frac{\partial \mu_{Zr}}{\partial C^0_{Zr}}
\end{equation}
We obtain the thermodynamic factor $\partial \mu_{Zr} / \partial
C^0_{Zr}$ using CVM-TO calculations. This factor is the same with
or without order corrections on triangles and tetrahedrons.
Therefore, the only difference these corrections induce on Zr
diffusion arises from $L_{ZrZr}$. In classical theory of
nucleation, the diffusion coefficient entering in the expression
\ref{condensation_rate} of the condensation rate  is only a
scaling factor for time and does not have any other influence on
kinetics. As a consequence the steady-state nucleation rate varies
linearly with Zr diffusion coefficient as it clearly appears when
combining equations \ref{condensation_rate} and
\ref{nucleation_rate_eq}. Thus small variations of $L_{ZrZr}$ with
the set of parameters used do not allow to explain the difference
of the nucleation rate: with order corrections, $L_{ZrZr}$ is far
from being half the value it is with only pair interactions (Fig.
\ref{Onsager_fig}). Thus slowdown of precipitation kinetics is not
due to a slowdown of Zr diffusion.

One possible explanation would be a difference of the interface
free energy $\sigma$. $J^{st}$ is really sensitive to this
parameter and one only needs a small decrease of $\sigma$ to
obtain a higher nucleation rate (see on fig. \ref{nuc_rate_fig}
the decrease of $J^{st}$ when $a^2\sigma$ is going from 57.0 to
64.1~meV). Such a decrease would explain too why precipitates have
a bigger size with only pair interactions. At $T=0$~K, we obtain
the same interface energy for all directions considered with the
two sets of parameters, but at finite temperature the
configurational entropy could lead to a difference of interface
free energy. Nevertheless, this needs to be confirmed, by CVM
calculations or using the Cahn-Hilliard method\cite{CAH58} for
instance. Another possible explanation to understand the different
kinetic pathways would be a different mobility of small clusters
with the two sets of parameters. But this would be quite
surprising, as we do not expect small clusters to be really mobile
because of the repulsion between vacancy and Zr solute atoms.


\section{CONCLUSIONS}

We built an atomistic kinetic model for Al-Zr binary system using
ab-initio calculations as well as experimental data. So as to be
as realistic as it should be at this atomic scale, this model
describes diffusion through vacancy jumps. Thanks to ab-initio
calculations we could improve usual thermodynamic descriptions
based on pair interactions and incorporate multisite interactions
for clusters containing more than two lattice points so as to
consider dependence of bonds with their local environment.

At temperatures lower than 1000~K, these energetic corrections due
to local order do not modify thermodynamics: the phase diagram
does not change when one does not consider these order
corrections. For higher temperatures they lead to a stabilization
of the ordered structure L1$_2$.

Concerning diffusion in the solid solution, these order
corrections on first nearest neighbor triangle and tetrahedron do
not really change the Onsager matrix, and thus diffusion
characteristics. They just lead to a slight slowdown of Zr
diffusion in the metastable solid solution. When looking at higher
temperatures, the slowdown of Zr diffusion is more important.

For precipitation, kinetics are slower with these interactions.
The slowdown is too important to be related to the small decrease
of Zr diffusion in the metastable solid solution at the same
temperature. One possibility would be a change of configurational
entropy contribution to interface free energy.

\begin{acknowledgments}
The authors are grateful to Prof. J.~M.~Sanchez for providing his
CVM program and for his invaluable help and advices on the
thermodynamics part, and to Dr F.~Soisson for very useful
discussions on Monte Carlo simulations, in particular for the
comparison with the classical theory of nucleation. They would
like to thank too Dr. B.~Legrand, Dr. G.~Martin, and Dr. C.~Sigli
for stimulating discussions. Financial support from Pechiney CRV
(France) is acknowledged.
\end{acknowledgments}

\appendix
\section{DETAILS OF AB-INITIO CALCULATIONS}
\label{appendix_abinitio} Ab initio calculations were carried out
using a full-potential linear-muffin-tin-orbital (FP-LMTO) method
\cite{AND75,MET88,MET89} in the version developed by Methfessel
and Van Schilfgaarde \cite{MET93}. The basis used contained 22
energy independent muffin-tin-orbitals (MTO) per Al and Zr site:
three $\kappa$ values for the orbitals s and p and two $\kappa$
values for the orbitals d where the corresponding kinetic energies
were $\kappa^2=0.01$ Ry (spd), $1.0$ Ry (spd), and $2.3$ Ry (sp).
A second panel with a basis composed of 22 energy independent MTO
with the same kinetic energies was used to make a correct
treatment of the 4p semicore states of Zr. The same uniform mesh
of points was used to make the integrations in the Brillouin zone
for valence and semicore states. The radii of the muffin-tin
spheres were chosen to have a compactness of 47.6\% for Al sites
and 54.1\% for Zr sites. Inside the muffin-tin spheres, the
potential is expanded in spherical harmonics up to $l=6$ and in
the interstitial region spherical Hankel functions of kinetic
energies $\kappa^2=1$ Ry and $3.0$ Ry were fitted up to $l=6$. The
calculations were performed in the generalized gradient
approximation (GGA) \cite{HOH64,KOH65} and the parameterization
used was the one of Perdew \etal \cite{PER96}.

\begin{footnotesize}

\begin{thebibliography}{10}

\bibitem{SOI00}
F.~Soisson and G.~Martin, ``{M}onte-{C}arlo simulations of the decomposition of
  metastable solid solutions: Transient and steady-state nucleation kinetics,''
  {\em Phys. Rev. B}, vol.~62, pp.~203--214, 2000.

\bibitem{LEB02}
Y.~{Le Bouar} and F.~Soisson, ``Kinetics pathway from embedded-atom-method
  potential: Influence of the activation barriers,'' {\em Phys. Rev. B},
  vol.~65, p.~0914103, 2002.

\bibitem{ATH00}
M.~Ath\`enes, P.~Bellon, and G.~Martin, ``Effects of atomic mobilities on phase
  separation kinetics: a {M}onte-{C}arlo study,'' {\em Acta mater.}, vol.~48,
  pp.~2675--2688, 2000.

\bibitem{ROU01}
J.~M. Roussel and P.~Bellon, ``Vacancy-assisted phase separation with
  asymmetric atomic mobility: Coarsening rates, precipitate composition, and
  morphology.,'' {\em Phys. Rev. B}, vol.~63, p.~184114, 2001.

\bibitem{DUCASTELLE}
F.~Ducastelle, {\em Order and Phase Stability in Alloys}.
\newblock North-Holland, Amsterdam, 1991.

\bibitem{LU91}
Z.~W. Lu, S.-H. Wei, A.~Zunger, S.~Frota-Pessoa, and L.~G. Ferreira,
  ``First-principles statistical mechanics of structural stability of
  intermetallic compounds,'' {\em Phys. Rev. B}, vol.~44, pp.~512--544, 1991.

\bibitem{CALPHAD}
N.~Saunders and A.~P. Miodownik, {\em {CALPHAD} -- Calculation of Phase
  Diagrams -- A Comprehensive Guide}.
\newblock Oxford: Pergamon, 1998.

\bibitem{DES97}
C.~Desgranges, F.~Deffort, S.~Poissonet, and G.~Martin, ``Interdiffusion in
  concentrated quaternary {A}g-{I}n-{C}d-{S}n alloys: Modelling and
  measurements,'' in {\em Defect and Diffusion Science Forum}, vol.~143--147,
  pp.~603--608, 1997.

\bibitem{DES98}
C.~Desgranges, {\em Compr\'ehension et Pr\'ediction du Comportement sous
  Irradiation Neutronique d'Alliages Absorbants \`a Base d'Argent}.
\newblock PhD thesis, Universit\'e Paris~XI Orsay, 1998.

\bibitem{MUL01}
S.~M\"uller, C.~Wolverton, L.-W. Wang, and A.~Zunger, ``Prediction of alloy
  precipitate shapes from first principles,'' {\em Europhysics Letters},
  vol.~55, pp.~33--39, 2001.

\bibitem{AST98}
M.~Asta, S.~M. Foiles, and A.~A. Quong, ``First-principles calculations of bulk
  and interfacial thermodynamic properties for fcc-based {A}l-{S}c alloys,''
  {\em Phys. Rev. B}, vol.~57, no.~18, pp.~11265--11275, 1998.

\bibitem{MUL02}
S.~M\"uller, L.-W. Wang, and A.~Zunger, ``First-principles kinetics theory of
  precipitate evolution in {A}l-{Z}n,'' {\em Modelling Simul. Mater. Sci.
  Eng.}, vol.~10, pp.~131--145, 2002.

\bibitem{RYU69}
N.~Ryum, ``Precipitation and recrystallization in an {A}l-0.5~wt.\%~{Z}r
  alloy,'' {\em Acta Metall.}, vol.~17, pp.~269--278, 1969.

\bibitem{ROB01}
J.~D. Robson and P.~B. Prangnell, ``Dispersoid precipitation and process
  modelling in zirconium containing commercial aluminium alloys,'' {\em Acta
  Mater.}, vol.~49, pp.~599--613, 2001.

\bibitem{NES72}
E.~Nes, ``Precipitation of the metastable cubic {A}l$_3${Z}r-phase in
  subperitectic {A}l-{Z}r alloys,'' {\em Acta Metall.}, vol.~20, pp.~499--506,
  1972.

\bibitem{PRO01}
L.~Proville and A.~Finel, ``Kinetics of the coherent order-disorder transion in
  {A}l$_3${Z}r,'' {\em Phys. Rev. B}, vol.~64, p.~054104, 2001.

\bibitem{AND75}
O.~K. Andersen, ``Linear methods in band theory,'' {\em Phys. Rev. B}, vol.~12,
  no.~8, pp.~3060--3083, 1975.

\bibitem{MET88}
M.~Methfessel, ``Elastic constants and phonon frequencies of {S}i calculated by
  a fast full-potential {LMTO} method,'' {\em Phys. Rev. B}, vol.~38, no.~2,
  pp.~1537--1540, 1988.

\bibitem{MET89}
M.~Methfessel, C.~O. Rodriguez, and O.~K. Andersen, ``Fast full-potential
  calculations with a converged basis of atom-centered linear muffin-tin
  orbitals: Structural and dynamic properties of silicon,'' {\em Phys. Rev. B},
  vol.~40, no.~3, pp.~2009--2012, 1989.

\bibitem{CLO02}
E.~Clouet, J.~M. Sanchez, and C.~Sigli, ``First-principles study of the
  solubility of {Z}r in {A}l,'' {\em Phys. Rev. B}, vol.~65, p.~094105, 2002.

\bibitem{JOM98}
G.~Jomard, L.~Magaud, and A.~Pasturel, ``Full-potential calculations using the
  generalized-gradient corrections: Structural properties of {T}i, {Z}r and
  {Hf} under compression,'' {\em Philos. Mag. B}, vol.~77, no.~1, pp.~67--74,
  1998.

\bibitem{MES93}
S.~V. Meshel and O.~J. Kleppa, ``Standard enthalpies of formation of 4d
  aluminides by direct synthesis calorimetry,'' {\em J. Alloys Compd.},
  vol.~191, pp.~111--116, 1993.

\bibitem{DES91}
P.~B. Desch, R.~B. Schwarz, and P.~Nash, ``Formation of metastable {L}1$_2$
  phases in {Al}$_3${Z}r and {A}l-12.5\%{X}-25\%{Z}r ({X}$\equiv${L}i, {C}r,
  {F}e, {N}i, {Cu}),'' {\em J. Less-Common Metals}, vol.~168, pp.~69--80, 1991.

\bibitem{AMA95}
C.~Amador, J.~J. Hoyt, B.~C. Chakoumakos, and D.~de~Fontaine, ``Theoretical and
  experimental study of relaxation in {A}l$_3${T}i and {A}l$_3${Z}r ordered
  phases,'' {\em Phys. Rev. Lett.}, vol.~74, no.~24, pp.~4955--4958, 1995.

\bibitem{SAN84}
J.~M. Sanchez, F.~Ducastelle, and D.~Gratias, ``Generalized cluster description
  of multicomponent systems,'' {\em Physica}, vol.~A 128, pp.~334--350, 1984.

\bibitem{LAK92}
D.~B. Laks, L.~G. Ferreira, S.~Froyen, and A.~Zunger, ``Efficient cluster
  expansion for substitutional systems,'' {\em Phys. Rev. B}, vol.~46, no.~19,
  pp.~12587--12605, 1992.

\bibitem{KIK51}
R.~Kikuchi, ``A theory of cooperative phenomena,'' {\em Phys. Rev.}, vol.~81,
  no.~6, pp.~988--1003, 1951.

\bibitem{SAN80}
J.~M. Sanchez and D.~{de Fontaine}, ``Ordering in fcc lattices with first- and
  second-neighbor interactions,'' {\em Phys. Rev. B}, vol.~21, p.~216, 1980.

\bibitem{MOH85}
T.~Mohri, J.~M. Sanchez, and D.~{de Fontaine}, ``Binary ordering prototype
  phase diagrams in the cluster variation approximation,'' {\em Acta Met.},
  vol.~33, pp.~1171--85, 1985.

\bibitem{LANDOLT25}
P.~Ehrhart, P.~Jung, H.~Schultz, and H.~Ullmaier, ``Atomic defects in metals,''
  in {\em {L}andolt-{B}\"ornstein, New Series, Group {III}} (H.~Ullmaier, ed.),
  vol.~25, Berlin: Springer-Verlag, 1991.

\bibitem{LEB99}
O.~{Le Bacq}, F.~Willaime, and A.~Pasturel, ``Unrelaxed vacancy formation
  energies in group-{IV} elements calculated by the full-potential linear
  muffin-tin orbital method: Invariance with crystal structure,'' {\em Phys.
  Rev. B}, vol.~59, pp.~8508--8515, 1999.

\bibitem{SIM77}
J.~P. Simon, ``{\'E}tude par trempe des interactions lacunes-impuret\'es dans
  les alliages dilu\'es {A}l-{Z}r et {A}l-{C}r,'' {\em Phys. Stat. Sol. (a)},
  vol.~41, p.~K107, 1977.

\bibitem{LANDOLT26}
H.~Bakker, H.~P. Bonzel, C.~M. Bruff, M.~A. Dayananda, W.~Gust, J.~Horvßth,
  I.~Kaur, G.~Kidson, A.~D. LeClaire, H.~Mehrer, G.~Murch, G.~Neumann,
  N.~Stolica, and N.~A. Stolwijk, ``Diffusion in solid metals and alloys,'' in
  {\em {L}andolt-{B}\"ornstein, New Series, Group {III}} (H.~Mehrer, ed.),
  vol.~26, Berlin: Springer-Verlag, 1990.

\bibitem{MAR73}
T.~Marumo, S.~Fujikawa, and K.~Hirano, ``Diffusion of zirconium in aluminum,''
  {\em Keikinzoku - J. Jpn. Inst. Light Met.}, vol.~23, p.~17, 1973.

\bibitem{BOC96}
J.~L. Bocquet, G.~Brebec, and Y.~Limoge, ``Diffusion in metals ans alloys,'' in
  {\em Physical Metallurgy} (R.~W. Cahn and P.~Haasen, eds.), ch.~7,
  pp.~536--668, Amsterdam: North-Holland, 1996.

\bibitem{PHILIBERT}
J.~Philibert, {\em Atom Movements - Diffusion and Mass Transport in Solids}.
\newblock Les Ulis, France: Les \'editions de physique, 1991.

\bibitem{ALLNATT}
A.~R. Allnatt and A.~B. Lidiard, {\em Atomic Transport in Solids}.
\newblock Cambridge University Press, 1993.

\bibitem{BOR00}
A.~Borgenstam, A.~Engstr\"om, L.~H\"oglund, and J.~Agren, ``{DICTRA}, a tool
  for simulation of diffusional transformations in alloys,'' {\em J. Phase
  Equil.}, vol.~21, pp.~269--280, 2000.

\bibitem{AND92}
J.-O. Andersson and J.~{\AA gren}, ``Models for numerical treatment of
  multicomponent diffusion in simple phases,'' {\em J. Appl. Phys.}, vol.~72,
  pp.~1350--1355, 1992.

\bibitem{CAM02}
C.~E. Campbell, W.~J. Boettinger, and U.~R. Kattner, ``Development of a
  diffusion mobility database for {N}i-base superalloys,'' {\em Acta Mat.},
  vol.~50, pp.~775--792, 2002.

\bibitem{MAR90}
G.~Martin, ``Atomic mobility in {C}ahn's diffusion model,'' {\em Phys. Rev. B},
  vol.~41, p.~2279, 1990.

\bibitem{NAS00}
M.~Nastar, V.~Y. Dobretsov, and G.~Martin, ``Self-consistent formulation of
  configurational kinetics close to equilibrium: the phenomenological
  coefficients for diffusion in crystalline solids,'' {\em Phil. Mag. A},
  vol.~80, p.~155, 2000.

\bibitem{ALL82}
A.~R. Allnatt, ``Einstein and linear response formulae for the phenomenological
  coefficients for isothermal matter transport in solids,'' {\em J. Phys. C:
  Solid State Phys.}, vol.~15, pp.~5605--5613, 1982.

\bibitem{MAR78}
G.~Martin, ``The theories of unmixing kinetics of solid solutions,'' in {\em
  Solid State Phase Transformation in Metals and Alloys}, (Orsay, France),
  pp.~337--406, Les \'Editions de Physique, 1978.

\bibitem{PORTER}
D.~A. Porter and K.~E. Easterling, {\em Phase Transformations in Metals and
  Alloys}.
\newblock London: Chapman \& Hall, 1992.

\bibitem{CAH58}
J.~W. Cahn and J.~Hilliard, ``Free energy of a nonuniform system. interface
  free energy,'' {\em J. Chem. Phys.}, vol.~28, pp.~258--267, 1958.

\bibitem{MET93}
M.~Methfessel and M.~van Schilfgaarde, ``Derivation of force theorem in
  density-functional theory: Application to the full-potential {LMTO} method,''
  {\em Phys. Rev. B}, vol.~48, no.~7, pp.~4937--4940, 1993.

\bibitem{HOH64}
P.~Hohenberg and W.~Kohn, ``Inhomogeneous electron gas,'' {\em Phys. Rev.},
  vol.~136, no.~3B, pp.~B864--B871, 1964.

\bibitem{KOH65}
W.~Kohn and L.~J. Sham, ``Self-consistent equations including exchange and
  correlations effects,'' {\em Phys. Rev.}, vol.~140, no.~4A, pp.~A1133--A1138,
  1965.

\bibitem{PER96}
J.~P. Perdew, K.~Burke, and M.~Ernzerhof, ``Generalized gradient approximation
  made simple,'' {\em Phys. Rev. Lett.}, vol.~77, no.~18, pp.~3865--3868, 1996.

\end{thebibliography}

\end{footnotesize}

\end{document}